\documentclass[%
 preprint,onecolumn,
 amsmath,amssymb,
 aps,
]{revtex4-1}
\usepackage{graphicx,color}
\usepackage{dcolumn}
\usepackage{bm}

\begin{document}


\title{Search for dark energy potentials in quintessence}
\author{Yusuke Muromachi}
\author{Akira Okabayashi}%
\altaffiliation{Graduate School of Human and Environmental Studies, Kyoto University, Kyoto 608-8501, Japan}%
\author{Daiki Okada}%
\affiliation{Department of Physics, Kyoto Sangyo University, Kyoto 603-8555, Japan}
\author{Tetsuya Hara} 
\email{hara@cc.kyoto-su.ac.jp}
\affiliation{Department of Physics, Kyoto Sangyo University, Kyoto 603-8555, Japan}
\author{Yutaka  Itoh}
\email{yitoh@cc.kyoto-su.ac.jp}
\affiliation{Department of Physics, Kyoto Sangyo University, Kyoto 603-8555, Japan}


\begin{abstract}
The time evolution of the equation of state $w$ for quintessence models with a scalar field as dark energy is studied up to the third derivative ($d^3w/da^3$) 
with respect to the scale factor $a$, in order to predict the future observations and specify the scalar potential parameters with the observables.
The third derivative of $w$ for general potential $V$ is derived and applied to several types of potentials.  
They are the inverse power-law ($V=M^{4+\alpha}/Q^{\alpha}$), the exponential ($V=M^4\exp{(\beta M/Q)}$), the mixed ( $V=M^{4+\gamma}\exp{(\beta M/Q)}/Q^{\gamma}$),
the cosine ($V=M^4(\cos (Q/f)+1)$) and the Gaussian types ($V=M^4\exp(-Q^2/\sigma^2)$), which are prototypical potentials for the freezing and thawing models.  
 If the parameter number for a potential form is $ n$, 
it is necessary to find at least for $n+2$ independent observations to identify the potential form and the evolution of the scalar field ($Q$ and $ \dot{Q} $). 
 Such observations would be the values of $ \Omega_Q, w, dw/da. \cdots $, and $ dw^n/da^n$. 
  From these specific potentials, we can predict the $ n+1 $ and higher derivative of $w$ ; $ dw^{n+1}/da^{n+1}, \cdots$. 
 Since four of the above mentioned potentials have two parameters, it is necessary to calculate the third derivative of $w$ for them to estimate the predict values.   
  If they are tested observationally, it will be understood whether the dark energy could be described by the scalar field with this potential.  
  At least it will satisfy the necessary conditions.  
  Numerical analysis for $d^3w/da^3$ are made under some specified parameters in the investigated potentials, except the mixed one.  
  It becomes possible to distinguish the potentials by the accurate observing $dw/da$ and $d^2w/da^2$ in some parameters.
\end{abstract}	

\maketitle

\section{Introduction}
The acceleration in the expansion of the universe was discovered by the intensive observations of the cosmology almost ten and several years ago \cite{1, 2}.  
Although the dark energy was introduced to cause the late-time accelerated universe, the physical mechanism and origin have been poorly understood \cite{3, 4}. 
Two theoretical viewpoints have been proposed so far.  One is associated with modification of gravity.  The other is associated with matter field theories \cite{5}.  
From the latter viewpoint, we explore the possibilities of the scalar fields in quintessence models and study how relevant to the dark energy.

  In the quintessence models, the scalar fields cause the time evolution of the universe.  Since the scalar field theories involve $ n$ independent 
  parameters, we notice that in principle $n$ time derivatives of the equation of state with observable $\Omega_Q$ and $w$ are enough to specify the scalar potentials
  and to predict the higher derivatives.  In this paper, we have carried out the calculations of the third derivative of the equation of state 
  for five scalar potentials to identify the models and to predict the future observations.  The parameters in the potentials can be determined 
  by the knowledge of the first, second,  and the higher derivatives, with the observable $\Omega_Q$ and $w$.  
  The first and second derivatives have been reported in the previous paper \cite{6}.

Usually, the variation of the equation of state $w$ for the dark energy is described by \cite{7,8,9,10} 
\vspace{-1.0cm}
\begin{align}
 w(a)=w_0+w_a(1-a) , \label{act0}
\end{align}
where  $a, w_0, $ and $ w_a$ are the scale factor ($a=1$ at current), the current value of $ w(a) $ and the first derivative of $ w(a)$ by $w_a=-dw/da$, respectively.

  We have extended the parameter space, in this paper, 
\begin{align}
w(a)=w_0+w_a(1-a)+\frac{1}{2}w_{a2}(1-a)^2 +\frac{1}{3!}w_{a3}(1-a)^3 , \label{act1}
\end{align}
where $w_{a2}=-d^2 w/da^2$ and $w_{a3}=-d^3 w/da^3$.
One of the new ingredients of this work in comparison with past works is the inclusion of this third derivative for the parameter space.

We follow the single scalar field formalism of Steinhardt {\it et al.} (1999) \cite{11, 12} and investigate three potentials 
for so-called freezing model \cite {13}, in which the field is rolling towards down its potential minimum,
as $ V=M^{4+\alpha}/Q^{\alpha}$  \cite{14}, $ V=M^4\exp(\beta M/Q)$, and $V=M^{4+\gamma}/Q^{\gamma}\exp (\zeta Q^2 /M_{pl}^2)$  \cite{15}. 
Two of them are supported by observational data \cite{16}.  
We study other two potentials for so-called thawing model, in which the field is nearly constant at first and then starts to evolve slowly down the potential;
 $V=M^4(\cos (Q/f)+1))$ \cite{17,18} and $V=M^4\exp(-Q^2/\sigma^2)$ \cite{18}. The cosine type is called the pseudo Nambu-Goldstone boson potential \cite{17,18}, which is the prototype potential of thawing model.  
The above mentioned potentials are motivated by particle physics.
Investigation of those potentials with the method \cite{6} is another main new ingredient of this work.  
Numerical analysis are made for $d^3w/da^3$ under some specified parameters in the investigated potentials, except the mixed one

 The goal of this paper is to explore the dark energy under the quintessence model in a single scalar field by assuming the potential. 
We have tried to increase the parameter space to examine the features of dark energy, by adding the third derivative.  
To determine the potential form we must observe the expansion history of the universe. 
If the parameter number is $ n$ for the potential form, it will be necessary for $n+2$ independent observations to determine 
the potential form, $Q$ and $\dot{Q}$ at some time for the time variation of scalar field.  
Such observations would be values of $ \Omega_Q, w, dw/da. \cdots $, and $ dw^n/da^n$. 
  From these specific potentials, we can predict the $ n+1 $ and higher derivative of $w$ ; $ dw^{n+1}/da^{n+1}, \cdots$. 
 Because four of the above mentioned potentials have two parameters, it is necessary to calculate the third derivative of $w$ for them 
 to estimate the predict values.   
 If they are the predicted one, it will be understood that the dark energy could be described by the scalar field with this potential.  
  At least it will satisfy the necessary conditions. One of the above mentioned potentials has three parameters, 
  so it is necessary to calculate the fourth derivative of $w$ to estimate the predict values, which is not calculated in this paper. 
   However, the principle would be the same to calculate them.

   In Sect. II, the equation of state for the scalar field are presented and 
the results of the first, second, and third derivatives of $w_Q$  are summarized, where the detailed calculations are displayed in Appendix.
 Three potentials of freezing model are studied in Sect. III, and two potentials of thawing model are described in Sect. IV.  
 The numerical analysis for the predicted $d^3w_Q/da^3$ are presented in Sect. V. 
 The conclusions and discussion are considered in Sect. VI.

\section{First, second, and third derivatives of $w_Q$}
 \subsection{Scalar field}
 For the dark energy, we consider the scalar field $Q(\bf{x},t)$, where the action for this field in the gravitational field is described by 
 \begin{align}
S=\int d^4 x \sqrt{-g} \left[ -\frac{1}{16 \pi G}R+ \frac{1}{2}g^{\mu \nu} \partial _ \mu Q \partial _ \nu Q -V( Q ) \right] +S_M ,  \label{act2}
\end{align}
 where $S_M$ is the action of the matter field and $G$ is the gravitational constant, occasionally putting $G=1$ \ \cite{4} .
 Neglecting the coordinate dependence, the equation for $Q(t)$ becomes 
 \begin{align}
\ddot{Q} + 3H \dot{Q}+V'=0  , \label{Qfield}
\end{align}
where $H$ is the Hubble parameter, overdot is the derivative with time, and $V'$ is the derivative with $Q$.  Putting $\kappa=8\pi/3$, $H$ satisfies the following equation
\begin{align}
H^2=\left(\frac{\dot{a}}{a} \right)^2=\kappa (\rho _B + \rho _ Q)=\kappa \rho_c  , \label{Friedmann}
\end{align}
where $\rho _B$, $\rho _ Q $ and $\rho_c$ are the energy density of the background, the scalar field, and the critical density of the universe.
The energy density and pressure for the scalar field are written by
\begin{align}
\rho _ Q =\frac{1}{2} \dot{Q}^2+V \label{density} ,   \end{align}
and
\begin{align}
p_ Q =\frac{1}{2} \dot{Q}^2-V  , \label{pressure}
\end{align}
respectively.  Then the parameter $w_Q$ for the equation of state is described by 
\begin{align}
w_ Q \equiv \frac{p_ Q}{\rho _ Q}=\frac{ \frac{1}{2} \dot{Q}^2-V}{ \frac{1}{2} \dot{Q}^2+V} . \label{state}
\end{align}

 \subsection{Time variation of $w_Q$}
 It is assumed that the current value of $w_Q$ is slightly different from a negative unity by $\Delta ( > 0)$ 
 \begin{align}
w_Q=-1+\Delta     .\label{w}
\end{align}
By using Eq. (\ref{state}), $\dot{Q}^2$ is written as 
\begin{align}
\dot{Q}^2=\frac{\Delta V}{1-\frac{\Delta}{2}} ,  \label{(2)}
\end{align}
which becomes, using the density parameter $\Omega_Q=\rho_Q/\rho_c$,  
 \begin{align}
\dot{Q}^2=2(\rho_c\Omega_Q-V) . \label{(3)}
\end{align}
Combining Eqs. (\ref{(2)}) and (\ref{(3)}), $V$ is given by
\begin{align}
V=\rho_c\Omega_Q\left(1-\frac{\Delta}{2}\right) . \label{potential}
\end{align}
From Eqs. (\ref{(3)}) and (\ref{potential}), $\dot{Q}$ is expressed 
\begin{align}
\dot{Q}=\sqrt{\Delta(\rho_c\Omega_Q)}. \label{13}
\end{align}

 Since $\rho_c$ is given by the observation through the Hubble parameter $H$, 
 $\dot{Q}$ is determined by  $\Omega_Q$ and $\Delta$, which also determine the value of $V$.  If we adopt the form and parameters of each potential, 
 the value of $V$ could be used to estimate the value of $Q$. 
 Actually, the evolution of $H$ in Eq. (\ref{Qfield}) depends on the background densities which include radiation density.  
 The effect of radiation density can be ignored in the near past ($z \leq 10^3$) and so is not considered in this work.

\subsection{First derivative of $w_Q$}
To investigate the variation of $w_Q$, we calculate $dw_Q/da$, using Eqs. (\ref{Qfield}), (\ref{density}) and (\ref{pressure}), after Ref. \cite{6}
\begin{align}
\frac{dw_Q}{da} & = \frac{1}{\dot{a}}\frac{d}{dt}\left(\frac{p_Q}{\rho_Q}\right)=\frac{1}{\dot{a}}\frac{\dot{p}_Q\rho_Q-p_Q\dot{\rho}_Q}{\rho^2_{Q}} \nonumber \\
& = \frac{2V\dot{Q}}{aH\rho^2_Q}\left(-3H\dot{Q}-\frac{V'}{V}\rho_Q\right) . \label{dw/da}
\end{align}
If the first derivative is observed, $\frac{V'}{V}M_{pl}$ is specified by, 
\begin{eqnarray}
\frac{V'}{V}M_{pl} = -\left(1-\frac{\Delta}{2}\right)^{-1}\sqrt{\frac{2\pi}{3\Delta\Omega_Q}}\left\{a\frac{dw_Q}{da}+6\Delta(1-\frac{\Delta}{2})\right\}.
\label{X1}
\end{eqnarray}
where $M_{pl}$ is the Planck mass.  
To investigate further, we must consider each potential form.

\subsection{Second derivative of $w_Q$}
From Eq. (\ref{dw/da}), the second derivative of $w_Q$  is given by
\begin{equation}
\frac{d^2w_Q}{da^2} = \frac{1}{\dot{a}^3\rho_Q^4}[(\ddot{p}_Q\rho_Q-p_Q\ddot{\rho}_Q)\dot{a}\rho^2_Q-(\dot{p_Q}\rho_Q-p_Q\dot{\rho}_Q)(\ddot{a}\rho_Q^2+2\dot{a}\rho_Q\dot{\rho}_Q)].   
\label{d^2w/da^2}
\end{equation}

After the calculation in the paper \cite{6}, $d^2w/da^2 $ becomes 
\begin{align}
 & \frac{d^2w_Q}{da^2}= \frac{3}{4\pi}\frac{\Omega_Q}{a^2}\left(1-\frac{\Delta}{2}\right) \nonumber \\ 
& \times \biggl[-\Delta M^2_{ \rm pl}\frac{V''}{V} +\sqrt{\frac{6\pi \Delta}{\Omega_Q}}\left(\left(1-\Delta\right)(6+\Omega_Q)-\frac{1}{3} \Omega_Q \right) \label{d^2w/da^2_A}  \\
& \quad \times M_{\rm pl}\left(\frac{V'}{V}\right) +\left(1-\frac{\Delta}{2}\right)M^2_{\rm pl}\left(\frac{V'}{V}\right)^2+\frac{8\pi\Delta}{\Omega_Q}(7-6\Delta) \biggr] . \nonumber 
\end{align}
In the limit of $\Delta \rightarrow 0$, the signature of $d^2w_Q/da^2$ is positive under the condition $V'/V \neq 0$.  
From this equation, we estimate $d^2w_Q/da^2$ for each potential in the following.
If the first derivative is observed, $\frac{V'}{V}M_{\rm pl}$ is estimated. Then $\frac{V''}{V}M_{\rm pl}$ could be estimated, if $\frac{d^2 w_Q}{da^2}$ is observed,  as
\begin{eqnarray}
\frac{V''}{V}M_{\rm pl}^2 = &-&\frac{1}{\Delta} \left[\left(1-\frac{\Delta}{2}\right)^{-1}\frac{4\pi a^2}{3\Omega_Q}\frac{d^2 w_Q}{da^2}-\sqrt{\frac{6\pi \Delta}{\Omega_Q}} \left \{ (1-\Delta)(6+\Omega_Q)-\frac{1}{3}\Omega_Q\right \} M_{\rm pl}\frac{V'}{V} \right. \nonumber \\
&-& \left. \left(1-\frac{\Delta}{2}\right)\left(M_{\rm pl}\frac{V'}{V}\right)^2 - \frac{8\pi}{\Omega_Q}\Delta(7-6\Delta) \right].
\label{Y2}
\end{eqnarray}

 \subsection{Third derivative of $w_Q$}
From Eq. (\ref{d^2w/da^2}), the third derivative of $w_Q$  is given by
\newline
\begin{eqnarray*}
\frac{d^3w_Q}{da^3}&=&\frac{1}{(\dot{a}\rho_Q)^5}\left[ \left \{(p_Q^{(3)}\rho_Q+\ddot{p_Q}\dot{\rho_Q}-\dot{p_Q}\ddot{\rho_Q}-p_Q\rho_Q^{(3)})\dot{a}\rho_Q^2 \right. \right. \\
&-& \left. \left. (\dot{p_Q}\rho_Q-p_Q\dot{\rho_Q})(a^{(3)}\rho_Q^2+4\ddot{a}\rho_Q\dot{\rho_Q}+2\dot{a}\dot{\rho_Q}^2+2\dot{a}\rho_Q\ddot{\rho_Q} ) \right \}\dot{a}\rho_Q \right. \\
&-& \left. \left \{(\ddot{p_Q}\rho_Q-p_Q\ddot{\rho_Q})\dot{a}\rho_Q^2-(\dot{p_Q}\rho_Q-p_Q\dot{\rho_Q})(\ddot{a}\rho_Q^2+2\dot{a}\rho_Q\dot{\rho_Q}) \right \}(3\ddot{a}\rho_Q+4\dot{a}\dot{\rho_Q})\right]
\end{eqnarray*}
\begin{eqnarray}
&=& \frac{3^2\left(1-\frac{\Delta}{2}\right)\Omega_Q}{2a^3(8\pi)^2}\left[-\sqrt{\frac{128\pi\Delta^3}{3}\Omega_Q}M_{\rm pl}^3\frac{V'''}{V}+16\pi\Delta \left \{10+3\Omega_Q\left(\frac{2}{3}-\Delta \right)-8\Delta\right \}M_{\rm pl}^2\frac{V''}{V} \right. \nonumber \\
&+& \left. \left( \sqrt{\frac{(8\pi)^3}{3}\Omega_Q\Delta} \left \{-14\Delta+9\Delta^2+\frac{164\Delta-104+\Omega_Q\left(\frac{64}{3}-46\Delta+45\Delta^2\right)}{\Omega_Q}\right \} \right. \right. \nonumber \\
&-& \left. \left. 9\left \{\Omega_Q\left(\frac{2}{3}-\Delta \right)-\frac{8}{3}\Delta\right \}\sqrt{\frac{(8\pi)^3\Delta}{3\Omega_Q}}\left \{ (1-\Delta)(6+\Omega_Q)-\frac{1}{3}\Omega_Q\right \} \right)M_{\rm pl}\frac{V'}{V} \right. \nonumber \\
&+& \left. 16\left(1-\frac{\Delta}{2}\right)\sqrt{\frac{8\pi}{3}\Omega_Q\Delta}M_{\rm pl}^3 \frac{V'}{V}\frac{V''}{V} \right. \nonumber \\
&+& \left. \left \{ 32\pi \left(1-\frac{\Delta}{2}\right)\left(\Omega_Q\Delta+6\Delta-3\right)-48\pi\left \{ \Omega_Q\left(\frac{2}{3}-\Delta \right)-\frac{8}{3}\Delta\right \} \left(1-\frac{\Delta}{2}\right)\right\}M_{\rm pl}^2\left(\frac{V'}{V}\right)^2 \right. \nonumber \\
&+& \left. \frac{128\pi^2\Delta}{3}\left(-99\Delta+54\Delta^2+\frac{42\Omega_Q-112+36\Delta^2+84\Delta}{\Omega_Q} \right) \right. \nonumber \\
&-& \left. \frac{384\pi^2\Delta}{\Omega_Q}(7-6\Delta)\left \{\Omega_Q\left(\frac{2}{3}-\Delta \right)-\frac{8}{3}\Delta\right \} \right].
\label{d3w/da3}
\end{eqnarray}
The detailed derivation of this equation is described in the Appendix.
\vspace{-0.1cm}
 
 In the limit of $\Delta \rightarrow 0$ and under the condition $\frac{V'}{V} \neq 0$, the signature of $\frac{d^3w_Q}{da^3}$  becomes negative.
From the above Eq. (\ref{d3w/da3}), $\frac{V'''}{V}M_{\rm pl}^3$ is given by
\begin{eqnarray}
\frac{V'''}{V}M_{\rm pl}^3 = &-& \sqrt{\frac{3}{128\pi \Delta^3\Omega_Q}}\left[\frac{d^3w_Q}{da^3}\frac{2a^3(8\pi)^2}{3^2\left(1-\frac{\Delta}{2}\right)\Omega_Q}-16\pi\Delta \left \{10+3\Omega_Q\left(\frac{2}{3}-\Delta \right)-8\Delta\right \}M_{\rm pl}^2\frac{V''}{V} \right. \nonumber \\
&-& \left. \left( \sqrt{\frac{(8\pi)^3}{3}\Omega_Q\Delta} \left \{-14\Delta+9\Delta^2+\frac{164\Delta-104+\Omega_Q\left(\frac{64}{3}-46\Delta+45\Delta^2\right)}{\Omega_Q}\right \} \right. \right. \nonumber \\
&-& \left. \left. 9\left \{\Omega_Q\left(\frac{2}{3}-\Delta \right)-\frac{8}{3}\Delta\right \}\sqrt{\frac{(8\pi)^3\Delta}{3\Omega_Q}}\left \{ (1-\Delta)(6+\Omega_Q)-\frac{1}{3}\Omega_Q\right \} \right)M_{\rm pl}\frac{V'}{V} \right. \nonumber \\
&-& \left. 16\left(1-\frac{\Delta}{2}\right)\sqrt{\frac{8\pi}{3}\Omega_Q\Delta}M_{\rm pl}\frac{V'}{V}M_{\rm pl}^2\frac{V''}{V} \right. \nonumber \\
&-& \left. 16\pi \left(1-\frac{\Delta}{2}\right)(5\Omega_Q\Delta+20\Delta-2\Omega_Q-6)\left(M_{\rm pl}\frac{V'}{V}\right)^2 \right. \nonumber \\
&-& \left. \frac{128\pi^2\Delta}{3}\left(-99\Delta+54\Delta^2+\frac{42\Omega_Q-112+36\Delta^2+84\Delta}{\Omega_Q} \right) \right. \nonumber \\
&+& \left. \frac{384\pi^2\Delta}{\Omega_Q}(7-6\Delta)\left \{\Omega_Q\left(\frac{2}{3}-\Delta \right)-\frac{8}{3}\Delta\right \} \right] .
\label{ThirdDerivative}
\end{eqnarray}
In the next section, we investigate the potential forms.  Although potential parts such as $V'/V, V''/V, and V'''/V$ are varying, 
the coefficients do not change in Eq. (\ref{d3w/da3}). 
Thus it is convenient to define the following notations,    
\begin{eqnarray}
A &=& \frac{3^2\left(1-\frac{\Delta}{2}\right)\Omega_Q}{2a^3(8\pi)^2} ,\nonumber \\
B &=& -\sqrt{\frac{128\pi\Delta^3}{3}\Omega_Q} ,\nonumber \\ 
C &=& 16\pi\Delta \left \{10+3\Omega_Q\left(\frac{2}{3}-\Delta \right)-8\Delta\right \} , \nonumber \\
D &=& \sqrt{\frac{(8\pi)^3}{3}\Omega_Q\Delta} \left \{-14\Delta+9\Delta^2+\frac{164\Delta-104+\Omega_Q\left(\frac{64}{3}-46\Delta+45\Delta^2\right)}{\Omega_Q}\right \} \nonumber \\ 
&-& 9\left \{\Omega_Q\left(\frac{2}{3}-\Delta \right)-\frac{8}{3}\Delta\right \}\sqrt{\frac{(8\pi)^3\Delta}{3\Omega_Q}}\left \{ (1-\Delta)(6+\Omega_Q)-\frac{1}{3}\Omega_Q\right \}, \nonumber \\
E &=& 16\left(1-\frac{\Delta}{2}\right)\sqrt{\frac{8\pi}{3}\Omega_Q\Delta}  ,\nonumber \\
F &=&  32\pi \left(1-\frac{\Delta}{2}\right)\left(\Omega_Q\Delta+6\Delta-3\right)-48\pi\left \{ \Omega_Q\left(\frac{2}{3}-\Delta \right)-\frac{8}{3}\Delta\right \} \left(1-\frac{\Delta}{2}\right),\nonumber \\
G &=& \frac{128\pi^2\Delta}{3}\left(-99\Delta+54\Delta^2+\frac{42\Omega_Q-112+36\Delta^2+84\Delta}{\Omega_Q} \right) \nonumber \\
&-& \frac{384\pi^2\Delta}{\Omega_Q}(7-6\Delta)\left \{\Omega_Q\left(\frac{2}{3}-\Delta \right)-\frac{8}{3}\Delta\right \} .
\label{sign}
\end{eqnarray}
By using these notations, Eqs. ({\ref{d3w/da3}}) and (\ref{ThirdDerivative}) become 
\begin{eqnarray}
\frac{d^3w_Q}{da^3} = A\left[B\frac{V'''}{V}M_{\rm pl}^3+C\frac{V''}{V}M_{\rm pl}^2+D\frac{V'}{V}M_{\rm pl}+E\frac{V'}{V}M_{\rm pl}\frac{V''}{V}M_{\rm pl}^2+F\left(\frac{V'}{V}M_{\rm pl}\right)^2+G\right] ,\nonumber \\
\label{WrittenThird}
\end{eqnarray}
and
\begin{eqnarray}
\frac{V'''}{V}M_{\rm pl}^3 &=& \frac{1}{B}\left[\frac{d^3w_Q}{da^3}\frac{1}{A}-C\frac{V''}{V}M_{\rm pl}^2-D\frac{V'}{V}M_{\rm pl}-E\frac{V'}{V}M_{\rm pl}\frac{V''}{V}M_{\rm pl}^2-F\left(\frac{V'}{V}M_{\rm pl}\right)^2-G\right] ,\nonumber \\
\end{eqnarray}
respectively.

 \section{Freezing model}
 In the freezing model, $w_Q$ will approach $-1$.  Then the first derivative of $w_Q$ is expected not positive ( $dw_Q/da  \leq 0 $). 
  If it is necessary,  we adopt the current scale factor is  $a=1$.
     In the following, we investigate the power inverse potential $V=M^4(M/Q)^{\alpha} \ (\alpha >0) $, 
 the exponential potential $V=M^4\exp(\beta M/Q) \  (\beta >0)$, and the mixed type potential $V=\frac{M^{4+\gamma}}{Q^{\gamma}}\exp (\zeta Q^2 /M_{\rm pl}^2) \  (\gamma, \zeta > 0) $,
  respectively.
 \subsection{$V=M^{4+\alpha}/Q^{\alpha}$}
 The parameters of this inverse power-law potential are $M$ and $\alpha$. 
From Eq. (\ref{potential}), $Q$ is given by
\begin{align}
 Q=\left(\frac{M^{4+\alpha}}{\rho_c\Omega_Q(1-\frac{\Delta}{2})}\right)^{1/\alpha} .\label{(4)}
\end{align}
If we take $Q=Q_0M_{\rm pl}$ at current, $M$ becomes
\begin{align} 
M= M_{\rm pl}\left(Q_0^{\alpha}\frac{\rho_c}{M_{\rm pl}^4}\Omega_Q (1-\frac{\Delta}{2})\right)^{1/(4+\alpha)} .
\end{align}
Then $Q_0, \Omega_Q, \Delta$, and $\alpha$ determine the parameter $M$, which means that parameters determining the accelerating behavior are $Q_0, \Omega_Q, \Delta$, and $\alpha$.
The problem is how to estimate $Q_0$ and $\alpha$.

\subsubsection {First derivative}
Using $V'/V=-\alpha/Q $, Eq. (\ref{dw/da}) becomes  
\begin{align}
\frac{dw_Q}{da} 
& = \frac{2V\dot{Q}}{aH Q\rho_Q}\left(\alpha-\frac{3H\dot{Q}Q}{\rho_Q}\right) . \label{4.14}
\end{align}
From Eq. (\ref{4.14}), $Q$ is derived as
 \begin{align}
 Q=\frac{2  \alpha  \rho_Q V \dot{Q}}{H(a\rho_Q^2\frac{dw_Q}{da}+6V\dot{Q}^2)},                                         \label{(6d)}
 \end{align}
 then $M_{\rm pl}/Q$ is given by 
 \begin{align}
 \frac{\alpha}{Q_0}= \left(1-\frac{\Delta}{2}\right)^{-1}\sqrt{\frac{2\pi}{3\Delta \Omega_Q}}\left(a\frac{dw_Q}{da}+6\Delta(1-\frac{\Delta}{2})\right).              \label{(6e)}
 \end{align}
 If $dw_Q/da$ is observed, $\alpha/Q_0$ will be determined by the observed values of $\Omega_Q, \Delta, $ and $dw_Q/da$.

\subsubsection {Second derivative}
Since the following relations are derived
\begin{align}
\frac{V''}{V}=\frac{\alpha(\alpha+1)}{Q^2}, \ \ \ 
\ \frac{V'}{V}=-\frac{\alpha}{Q}, \ \ \  \ \left(\frac{V'}{V}\right)^2=\frac{\alpha^2}{Q^2}, 
\label{V2Valpha}
\end{align}
we substitute them into Eq. (\ref{d^2w/da^2_A}) and obtain
\begin{align}
\frac{d^2w_Q}{da^2} & = \frac{3}{4\pi}\frac{\Omega_Q}{a^2}\left(1-\frac{\Delta}{2}\right)
 \times \biggl[-\Delta \frac{\alpha(\alpha+1)}{(Q/M_{\rm pl})^2}-\sqrt{\frac{6\pi \Delta}{\Omega_Q}}\left(\left(1-\Delta\right)(6+\Omega_Q)-\frac{1}{3}\Omega_Q\right)\left(\frac{\alpha}{Q/M_{\rm pl}}\right)  \label{4.57} \\
& \quad  +\left(1-\frac{\Delta}{2}\right)\left(\frac{\alpha}{Q/M_{\rm pl}}\right)^2+\frac{8\pi\Delta}{\Omega_Q}(7-6\Delta) \biggr] .  \nonumber 
\end{align}
If $dw_Q/da$ is observed, $(Q/M_{\rm pl})/\alpha$ will be determined by Eq. (\ref{(6e)}).  If $d^2w_Q/da^2$ is observed, one could estimate the value of $\alpha$ from the above equation as
\begin{align}
\alpha & = -\left(\frac{Q/M_{\rm pl}}{\alpha}\right)^2 \biggl[\left(1-\frac{\Delta}{2}\right)^{-1}\frac{4\pi}{3}\frac{a^2}{\Omega_Q}\frac{d^2w_Q}{da^2} 
+\sqrt{\frac{6\pi \Delta}{\Omega_Q}}\left(\left(1-\Delta\right)(6+\Omega_Q)-\frac{1}{3}\Omega_Q\right)\left(\frac{\alpha}{Q/M_{\rm pl}}\right)  \nonumber  \\
& -\left(1-\frac{3\Delta}{2}\right)\left(\frac{Q/M_{\rm pl}}{\alpha}\right)^{-2}- \frac{8\pi\Delta}{\Omega_Q}(7-6\Delta) \biggr]^{-1} \times \Delta.  \label{4.57a}
\end{align}

\subsubsection {Third derivative}
Since the following equations are derived
\begin{eqnarray}
\frac{V'''}{V}=-\frac{\alpha(\alpha+1)(\alpha+2)}{Q^3}, \ \ \ \ \frac{V'}{V}\frac{V''}{V}=-\frac{\alpha^2(\alpha+1)}{Q^3} ,
\end{eqnarray}
we substitute them into Eq. (\ref{WrittenThird}), using Eq. (\ref{sign}), and obtain
\begin{eqnarray}
\frac{d^3w_Q}{da^3} &=& A\left[-B\alpha(\alpha+1)(\alpha+2)\left(\frac{M_{\rm pl}}{Q}\right)^3 +C\alpha(\alpha+1)\left(\frac{M_{\rm pl}}{Q}\right)^2 -D\alpha\left(\frac{M_{\rm pl}}{Q}\right) \right. \nonumber \\
&-& \left. E\alpha^2(\alpha+1)\left(\frac{M_{\rm pl}}{Q}\right)^3  +F\alpha^2\left(\frac{M_{\rm pl}}{Q}\right)^2 +G \right]  .
\end{eqnarray}
Because we get $Q, \alpha$ through the observations of $dw_Q/da, d^2w_Q/da^2$, we can predict the third derivative for this potential.

\subsection{$V=M^4\exp(\beta M/Q)$ }
This exponential type potential has also two independent parameters of  $\beta$ and $M$. From Eq. (\ref {potential}), the potential relates to the observables
\begin{eqnarray}
 V=\rho_c \Omega_Q \left(1-\frac{\Delta}{2}\right)  = M^4 \exp\left(\frac{\beta M}{Q}\right) , \nonumber \\
\end{eqnarray}
which is written by
\begin{eqnarray}
\frac{\beta M}{Q} = \ln \left[\frac{\rho_c \Omega_Q \left(1-\frac{\Delta}{2}\right)}{M^4}\right] .
\label{betaM/Q}
\end{eqnarray}

\subsubsection{First derivative}
Since the first derivative of the potential is $ V' = M^4\exp\left(\frac{\beta M}{Q}\right)\left(-\frac{\beta M}{Q^2}\right)$, then 
\begin{eqnarray}
\frac{V'}{V} = -\frac{\beta M}{Q^2}.
\label{V'/V}
\end{eqnarray}
Using Eqs. (\ref{dw/da}), (\ref{Friedmann}), and (\ref{13}), the first derivative of $w_Q$ becomes
\begin{eqnarray}
\frac{dw_Q}{da} &=& \frac{2V}{a\rho_Q}\left(\frac{\beta M}{Q^2}M_{\rm pl}\sqrt{\frac{3\Delta\Omega_Q}{8\pi}}-3\Delta\right) \nonumber .\\
\end{eqnarray}
We get $\frac{\beta M}{Q^2}M_{\rm pl}$ from the observables 
\begin{eqnarray}
\frac{\beta M}{Q} \frac{M_{\rm pl}}{Q} = \left(1-\frac{\Delta}{2}\right)\sqrt{\frac{2\pi}{3\Delta\Omega_Q}}\left(a\frac{dw_Q}{da}+6\Delta(1-\frac{\Delta}{2})\right). \nonumber \\
\label{FirstDerivativeExp}
\end{eqnarray}
In the following Eq. (\ref{Mpl/Q}), we can estimate $M_{\rm pl}/Q$ by the observables.  After then we can estimate $\beta M/Q$ by the observables through this equation.

\subsubsection{Second derivative}
Since the second derivative of the potential is 
\begin{eqnarray}
V'' = M^4\exp\left(\frac{\beta M}{Q}\right)\left(-\frac{\beta M}{Q^2}\right)^2+M^4\exp\left(\frac{\beta M}{Q}\right)\frac{2\beta M}{Q^3} ,\nonumber \\
\end{eqnarray}
it is derived
\begin{eqnarray}
\frac{V''}{V} = \left(-\frac{\beta M}{Q^2}\right)^2+\frac{2\beta M}{Q^3}.
\label{V''/V}
\end{eqnarray}
The second derivative of $w_Q$ is obtained by Eq. (\ref{d^2w/da^2_A})
\begin{eqnarray}
\frac{d^2 w_Q}{da^2} &=& \frac{3}{16\pi}\frac{\Omega_Q}{a^2} \left(1-\frac{\Delta}{2}\right) \left[-4\Delta \left\{\left(\frac{\beta M}{Q}\right)^2\left(\frac{M_{\rm pl}}{Q}\right)^2+2\left(\frac{\beta M}{Q}\right)\left(\frac{M_{\rm pl}}{Q}\right)^2 \right\} \right. \nonumber \\
&-& \left. 6\sqrt{\frac{8\pi \Delta}{3\Omega_Q}} \left \{ (1-\Delta)(6+\Omega_Q)-\frac{1}{3}\Omega_Q\right \} \left(\frac{\beta M}{Q}\right)\left(\frac{M_{\rm pl}}{Q}\right) \right. \nonumber \\
&+& \left. 4 \left(1-\frac{\Delta}{2}\right)\left\{\left(\frac{\beta M}{Q}\right)\left(\frac{M_{\rm pl}}{Q}\right)\right\}^2 + \frac{32\pi}{\Omega_Q}\Delta(7-6\Delta) \right]  . 
\end{eqnarray}
From the observation of $\frac{dw_Q}{da}$, it is derived the value $\frac{\beta M}{Q}\frac{M_{\rm pl}}{Q}$ in Eq. (\ref {FirstDerivativeExp}). 
Then we rewrite the above equation by
\begin{eqnarray}
\frac{M_{\rm pl}}{Q} = &-&\frac{1}{2\Delta}\left(\frac{\beta M}{Q}\frac{M_{\rm pl}}{Q}\right)^{-1}\left[ \left\{\left(1-\frac{\Delta}{2}\right)^{-1}\frac{4\pi a^2}{3\Omega_Q}\frac{d^2 w_Q}{da^2} \right.\right. \nonumber \\
&+& \left.\left. \sqrt{\frac{6\pi \Delta}{\Omega_Q}} \left \{ (1-\Delta)(6+\Omega_Q)-\frac{1}{3}\Omega_Q\right \} \left(\frac{\beta M}{Q}\frac{M_{\rm pl}}{Q}\right) \right.\right. \nonumber \\
&-& \left.\left. \left(1-\frac{3\Delta}{2}\right)\left(\frac{\beta M}{Q}\frac{M_{\rm pl}}{Q}\right)^2 - \frac{8\pi}{\Omega_Q}\Delta(7-6\Delta) \right\}\right].
\label{Mpl/Q}
\end{eqnarray}
Then we find out $\frac{M_{\rm pl}}{Q}$.  From Eq. (\ref{FirstDerivativeExp}), $\frac{\beta M}{Q}$ is estimated and $M$ is derived through Eq. (\ref{betaM/Q}) by
\begin{eqnarray}
M = \left[\rho_c \Omega_Q \left(1-\frac{\Delta}{2}\right)\exp\left(-\frac{\beta M}{Q}\right)\right]^{\frac{1}{4}}.
\end{eqnarray}
The value of $\beta$ is estimated through Eq. (\ref{betaM/Q}).
  As the two parameters of $\beta$ and $M$ are specified, it becomes possible to predict the third derivative of $w$.

\subsubsection{Third derivative}
Since the third derivative of the potential is 
\begin{eqnarray}
V''' &=& M^4\exp\left(\frac{\beta M}{Q}\right)\left(-\frac{\beta M}{Q^2}\right)^3+4M^4\exp\left(\frac{\beta M}{Q}\right)\left(-\frac{\beta^2 M^2}{Q^5}\right) \nonumber \\
&+& M^4\exp\left(\frac{\beta M}{Q}\right)\left(-\frac{\beta M}{Q^2}\right)\frac{2\beta M}{Q^3}+M^4\exp\left(\frac{\beta M}{Q}\right)\left(-\frac{6\beta M}{Q^4}\right) , 
\end{eqnarray}
$V'''/V$ leads to 
\begin{eqnarray}
\frac{V'''}{V} &=& -\frac{\beta^3 M^3}{Q^6}-\frac{6\beta^2 M^2}{Q^5}-\frac{6\beta M}{Q^4}.
\end{eqnarray}
Then the third derivative of $w$ is given through Eqs. (\ref{WrittenThird}) and (\ref{sign}) by
\begin{eqnarray}
\frac{d^3w_Q}{da^3}&=& A \left[-B\left\{\left(\frac{\beta M}{Q}\frac{M_{\rm pl}}{Q}\right)^3+6\left(\frac{\beta M}{Q}\right)^2\left(\frac{M_{\rm pl}}{Q}\right)^3+6\left(\frac{\beta M}{Q}\right)\left(\frac{M_{\rm pl}}{Q}\right)^3 \right\} \right. \nonumber \\
&+& \left. C\left\{\left(\frac{\beta M}{Q}\frac{M_{\rm pl}}{Q}\right)^2+2\left(\frac{\beta M}{Q}\right)\left(\frac{M_{\rm pl}}{Q}\right)^2\right\} -D\left(\frac{\beta M}{Q}\frac{M_{\rm pl}}{Q}\right) \right. \nonumber \\
&-& \left. E\left\{\left(\frac{\beta M}{Q}\frac{M_{\rm pl}}{Q}\right)^3+2\left(\frac{\beta M}{Q}\right)^2\left(\frac{M_{\rm pl}}{Q}\right)^3\right\}+ F \left(\frac{\beta M}{Q}\frac{M_{\rm pl}}{Q}\right)^2+G \right] .
\end{eqnarray}
This is the predictable value for this potential.

\subsection{$V= \frac{M^{4+\gamma}}{Q^{\gamma}} \exp\left(\frac{\zeta Q^2}{M_{\rm pl}^2}\right)$}
There are three parameters $\zeta, M, $ and $\gamma$ for this mixed type potential.  If we use the relation  of Eq. (\ref{potential}) for the potential with the observables, the parameter $M$ is expressed by  
\begin{eqnarray}
M = \left[Q^{\gamma}\rho_c \Omega_Q \left(1-\frac{\Delta}{2}\right)\exp\left(-\frac{\zeta Q^2}{M_{\rm pl}^2}\right)\right]^{\frac{1}{4+\gamma}},
\label{parametaM}
\end{eqnarray}
where there are three unspecified  $\gamma, \zeta$, and $Q$ parameters.

\subsubsection{First derivative}
Since the following relations are derived,
\begin{eqnarray}
V' &=& -\frac{\gamma M^{4+\gamma}}{Q^{\gamma+1}}\exp\left(\frac{\zeta Q^2}{M_{\rm pl}^2}\right)+\frac{M^{4+\gamma}}{Q^{\gamma}}\exp\left(\frac{\zeta Q^2}{M_{\rm pl}^2}\right)\frac{2\zeta Q}{M_{\rm pl}^2} , \nonumber \\
\frac{V'}{V} &=& -\frac{\gamma}{Q}+\frac{2\zeta Q}{M_{\rm pl}^2} ,
\end{eqnarray}
we substitute them into Eq. (\ref{X1}) 
\begin{eqnarray}
-\gamma \left(\frac{M_{\rm pl}}{Q}\right)+2\zeta \left(\frac{Q}{M_{\rm pl}}\right) = -\left(1-\frac{\Delta}{2}\right)^{-1}\sqrt{\frac{2\pi}{3\Delta \Omega_Q}}\left(a\frac{dw_Q}{da}+6\Delta(1-\frac{\Delta}{2})\right) .
\label{3parameta}
\end{eqnarray}
From the observables, including $\Delta, \Omega_Q$, and $dw_Q/da$, we can estimate 
\begin{eqnarray}
\frac{V'}{V}M_{\rm pl} = -\gamma \left(\frac{M_{\rm pl}}{Q}\right)+2\zeta \left(\frac{Q}{M_{\rm pl}}\right) =X, 
\label{48}
\end{eqnarray}
where we  put $X= \frac{V'}{V}M_{\rm pl}$.

\subsubsection{Second derivative}
Since the following equations are obtained
\begin{eqnarray}
V'' &=& \frac{M^{4+\gamma}}{Q^{\gamma}}\exp\left(\frac{\zeta Q^2}{M_{\rm pl}^2}\right)\left[\frac{\gamma(\gamma+1)}{Q^2}-\frac{2\zeta(2\gamma-1)}{M_{\rm pl}^2}+\left(\frac{2\zeta Q}{M_{\rm pl}^2}\right)^2\right] , \\
\frac{V''}{V} &=& \frac{\gamma(\gamma+1)}{Q^2}-\frac{2\zeta(2\gamma-1)}{M_{\rm pl}^2}+\left(\frac{2\zeta Q}{M_{\rm pl}^2}\right)^2 ,
\end{eqnarray}
we substitute them into Eq. (\ref{Y2}) and derive for $M_{\rm pl}^2\frac{V''}{V}=Y$ ;
\[ Y=\gamma(\gamma+1)\left(\frac{M_{\rm pl}}{Q}\right)^2-2\zeta(2\gamma-1)+\left(2\zeta \frac{Q}{M_{\rm pl}}\right)^2 =- \frac{1}{\Delta} \biggl[ \frac{4\pi a^2}{3\Omega_Q\left(1-\frac{\Delta}{2}\right)}\frac{d^2w_Q}{da^2} \nonumber \] 
\[- \sqrt{\frac{6\pi \Delta}{\Omega_Q} } \left\{ (1-\Delta)(6+\Omega_Q)-\frac{1}{3}\Omega_Q \right\} X - \left(1-\frac{\Delta}{2}\right) X^2- \frac{8\pi}{\Omega_Q}\Delta(7-6\Delta) \biggr].
\label{ポテンシャルから導かれるV''/V}
\]
From the observables, including $dw^2_Q/da^2$, we can estimate $Y$.  If we make the square of $X$
\begin{eqnarray*}
X^2 = \gamma^2\left(\frac{M_{\rm pl}}{Q}\right)^2-4\gamma\zeta+\left(2\zeta \frac{Q}{M_{\rm pl}}\right)^2,
\end{eqnarray*}
$Y$ is expressed by
\begin{equation}
Y=\gamma(\gamma+1)\left(\frac{M_{\rm pl}}{Q}\right)^2-2\zeta(2\gamma-1)+\left(2\zeta \frac{Q}{M_{\rm pl}}\right)^2 = X^2+\gamma \left(\frac{M_{\rm pl}}{Q} \right)^2 + 2 \zeta.
\label{Ytozeta}
\end{equation}
Then we can estimate $\gamma\left(\frac{M_{\rm pl}}{Q}\right)^2+2\zeta$ from $X$ and $Y$.

\subsubsection{Third derivative}
 There is still unspecified parameter, which is different from the potentials with two parameters.
Checking the third derivative of $w_Q$ in Eq. (\ref{d3w/da3}), there is still unknown term $M_{\rm pl}^3\frac{V'''}{V}$, which must be investigated.  The third derivative of the potential is  
\begin{eqnarray}
V''' &=& \frac{M^{4+\gamma}}{Q^{\gamma}}\exp\left(\frac{\zeta Q^2}{M_{\rm pl}^2}\right)\left[-\frac{\gamma(\gamma+1)(\gamma+2)}{Q^3}
+\frac{3\gamma^2}{Q}\frac{2\zeta}{M_{\rm pl}^2}+3Q(1-\gamma)\left(\frac{2\zeta}{M_{\rm pl}^2}\right)^2+\left(\frac{2\zeta Q}{M_{\rm pl}^2}\right)^3\right] , \nonumber \\
\end{eqnarray}
then $V'''/V$ is given by
\begin{eqnarray}
\frac{V'''}{V} = -\frac{\gamma(\gamma+1)(\gamma+2)}{Q^3}+\frac{3\gamma^2}{Q}\frac{2\zeta}{M_{\rm pl}^2}+3Q(1-\gamma)\left(\frac{2\zeta}{M_{\rm pl}^2}\right)^2+\left(\frac{2\zeta Q}{M_{\rm pl}^2}\right)^3 .
\end{eqnarray}
If we use the third power of $X$, $Z=M_{\rm pl}^3\frac{V'''}{V}$ is expressed as
\begin{eqnarray}
Z=M_{\rm pl}^3\frac{V'''}{V} &=& X^3 -3\gamma^2\left(\frac{M_{\rm pl}}{Q}\right)^3-2\gamma\left(\frac{M_{\rm pl}}{Q}\right)^3+12\zeta^2\left(\frac{Q}{M_{\rm pl}}\right) \nonumber    \\
&=& -2X^3 +3XY-2\gamma\left(\frac{M_{\rm pl}}{Q}\right)^3 ,
\label{ポテンシャルから導かれるV'''/V}
\end{eqnarray}
where we have used $XY=X^3-\gamma^2\left(\frac{M_{\rm pl}}{Q}\right)^3+4\zeta^2\left(\frac{Q}{M_{\rm pl}}\right)$.
If $dw_Q^3/da^3$ is observed, $Z$ could be estimated from Eq. (\ref{d3w/da3}).  
 So it is possible to specify three parameters $ \gamma, \zeta$, and $Q/M_{\rm pl}$ from the observables $X, Y,$ and $Z$.

 From Eq. (\ref{Ytozeta}), we put $2\zeta$ into Eq. (\ref{48}) and obtain 
 \begin{eqnarray}
 -2\gamma \left(\frac{M_{\rm pl}}{Q}\right)=X-(Y-X^2)\left(\frac{Q}{M_{\rm pl}}\right),  \nonumber
 \end{eqnarray}
 and put it into Eq. (\ref{ポテンシャルから導かれるV'''/V}).  Then we get
 \begin{eqnarray}
 (2X^3-3XY+Z)\left(\frac{Q}{M_{\rm pl}}\right)^2+(Y-X^2)\left(\frac{Q}{M_{\rm pl}}\right)-X=0.
 \end{eqnarray}
  Because $Q/M_{\rm pl}$ is derived from the above equation as
 \[ \frac{Q}{M_{\rm pl}}=\frac{X^2-Y+\sqrt{9X^4-14X^2Y+4XZ+Y^2 }}{2(2X^3-3XY+Z)}, \]
$\gamma$ is estimated from Eq. (\ref {ポテンシャルから導かれるV'''/V}), $\zeta$ is derived through Eq. (\ref {Ytozeta}), and $M$ is estimated by Eq. (\ref{parametaM}), respectively. 
For this potential, three parameters are specified through the observations $dw_Q/da, d^2w_Q/da^2,$ and $d^3w_Q/da^3$.  
However, it is necessary to calculate the fourth derivative of the potential to predict $d^4w_Q/da^4$.

\section{Thawing Model}
The definition of the thawing model is that the equation of state is $w=-1$ at early times and then it increases from $-1$, so it is expected  $\frac{dw_Q}{da} \geq  0$ .

\subsection{$V=M^4\left[\cos\left(\frac{Q}{f}\right)+1\right]$ }
For this cosine type potential, there are two parameters $M$ and $f$, where $f \  (>0) $ is the energy scale of spontaneous symmetry break down.  The potential is related to the observation by Eq. (\ref{potential}) as 
\begin{eqnarray}
\rho_c \Omega_Q \left(1-\frac{\Delta}{2}\right) = M^4\left[\cos\left(\frac{Q}{f}\right)+1\right] \left(= 2M^4\cos^2\left(\frac{Q}{2f}\right)\right).
\label{THW}
\end{eqnarray}

\subsubsection{First derivative}
Since the first derivative of the potential is $V' = -\frac{M^4}{f}\sin\left(\frac{Q}{f}\right)$, then
\begin{eqnarray}
X=\frac{V'}{V} = -\frac{\sin\left(\frac{Q}{f}\right)}{f\left[\cos\left(\frac{Q}{f}\right)+1\right]} =-\frac{1}{f} \tan \left(\frac{Q}{2f}\right) ,
\label{PN1}
\end{eqnarray}
where we put $X=V'/V$.  From Eq. (\ref{dw/da}), the first derivative of $w_Q$ becomes 
\begin{eqnarray}
\frac{dw_Q}{da} = \frac{2-\Delta}{a}\left(-3\Delta-X\sqrt{\frac{3\Delta\Omega_Q}{8\pi }}\right) . \nonumber 
\end{eqnarray}
If $dw_Q/da$ is observed, $X$ is estimated from
\begin{eqnarray}
X = -\left(1-\frac{\Delta}{2}\right)^{-1}\sqrt{\frac{2\pi}{3\Delta\Omega_Q}}\left(a\frac{dw_Q}{da}+6\Delta(1-\frac{\Delta}{2})\right) . 
\label{COSX}
\end{eqnarray}
If $dw_Q/da \geq 0$, then $ X < 0 $. It means 
\[\frac{1}{f} \tan \left(\frac{Q}{2f}\right)  > 0.\]

\subsubsection{Second derivative}
Since the second derivative of the potential is $V'' = -\frac{M^4}{f^2}\cos\left(\frac{Q}{f}\right)$, $Y=V''/V$ is given by
\begin{eqnarray}
Y=\frac{V''}{V} = -\frac{\cos\left(\frac{Q}{f}\right)}{f^2\left[\cos\left(\frac{Q}{f}\right)+1\right]}.
\label{PNGBポテンシャルの組み合わせ2}
\end{eqnarray}
The second derivative of $w_Q$ is derived by Eq. (\ref{d^2w/da^2_A}) as
\begin{eqnarray}
\frac{d^2 w_Q}{da^2} &=& \frac{3}{16\pi}\frac{\Omega_Q}{a^2} \left(1-\frac{\Delta}{2}\right) \left[-4\Delta Y +6\sqrt{\frac{8\pi G \Delta}{3\Omega_Q}} \left \{ (1-\Delta)(6+\Omega_Q)-\frac{1}{3}\Omega_Q\right \} X \right. \\
&+& \left. 4 \left(1-\frac{\Delta}{2}\right) X^2 + \frac{32\pi }{\Omega_Q}\Delta(7-6\Delta) \right] . \nonumber 
\end{eqnarray}
If $d^2w_Q/da^2$ is observed, it becomes possible to estimate $Y$  as
\begin{eqnarray}
Y &=&  -\frac{1}{\Delta} \left[\left(1-\frac{\Delta}{2}\right)^{-1}\frac{4\pi a^2}{3\Omega_Q}\frac{d^2 w_Q}{da^2}-\sqrt{\frac{6\pi\Delta}{\Omega_Q}} \left \{ (1-\Delta)(6+\Omega_Q)-\frac{1}{3}\Omega_Q\right \} X \right. \nonumber \\
&-& \left. \left(1-\frac{\Delta}{2}\right) X^2 - \frac{8\pi}{\Omega_Q}\Delta(7-6\Delta) \right] .
\label{COSY}
\end{eqnarray}
From Eqs. (\ref{PN1}) and (\ref{PNGBポテンシャルの組み合わせ2}), $ f$ is estimated by
\begin{eqnarray}
f = \frac{1}{\sqrt{X^2-2Y}}.
\label{sqrt}
\end{eqnarray}
From Eq. (\ref{PN1}), $Q$ is estimated from $X/Y=f\tan (Q/f)$ as
\begin{eqnarray}
Q = \frac{1}{\sqrt{X^2-2Y}}\tan^{-1}\left(\frac{X}{Y}\sqrt{X^2-2Y}\right),
\label{YNEG}
\end{eqnarray}
which is equivalent, from $X=-\frac{1}{f}\tan ^{-1}(Q/2f)$, to
\begin{eqnarray}
Q = \frac{2}{\sqrt{X^2-2Y}}\tan^{-1}\left(-\frac{X}{\sqrt{X^2-2Y}}\right).
\end{eqnarray}
From Eq. (\ref{THW}), $M$ is also determined.  Then it becomes possible to predict the third derivative. 
It must be noted from Eq. (\ref{YNEG}) that $Y$ is negative.

\subsubsection{Third derivative}
Since the following relations are derived
\begin{eqnarray}
V''' &=& \frac{M^4}{f^3}\sin\left(\frac{Q}{f}\right) ,\\
\frac{V'''}{V} &=& \frac{\sin\left(\frac{Q}{f}\right)}{f^3\left[\cos\left(\frac{Q}{f}\right)+1\right]} = -\frac{X}{f^2} = -X(X^2-2Y),
\end{eqnarray}
the third derivative of $w_Q$ is given through Eq. (\ref{WrittenThird}) by 
\begin{eqnarray}
\frac{d^3w_Q}{da^3} &=& A\left[BX(2Y-X^2)M_{\rm pl}^3+CYM_{\rm pl}^2 + DXM_{\rm pl} +EXY M_{\rm pl}^3+ FX^2M_{\rm pl}^2 +G \right].
\end{eqnarray}
This is the predictable value for this potential.

\subsection{$V=M^4\exp\left(-\frac{Q^2}{\sigma^2}\right)$}
For this Gaussian type potential, there are two parameters which are $M$ and $\sigma$.  

\subsubsection{First Derivative of $w_Q$}
Since the first derivative of the potential is $V'=-\frac{2Q}{\sigma^2}M^4\exp\left(-\frac{Q^2}{\sigma^2}\right)$, $V'/V$ becomes 
\begin{eqnarray}
\frac{V'}{V}=-\frac{2Q}{\sigma^2}.
\label{SIGV'/V}
\end{eqnarray}
From Eq. (\ref{X1}),
\begin{eqnarray}
\frac{2Q}{\sigma^2} = \left(1-\frac{\Delta}{2}\right)^{-1}\sqrt{\frac{2\pi}{3\Delta\Omega_Q}}\left(a\frac{dw_Q}{da}+6\Delta(1-\frac{\Delta}{2})\right) .
\label{Q/sigma2}
\end{eqnarray}
If $dw_Q/da$ is observed, $\frac{2Q}{\sigma^2}$ could be estimated.

\subsubsection{Second derivative}
Since the second derivative of the potential is $V'' = \left(-\frac{2}{\sigma^2}+\frac{4Q^2}{\sigma^4}\right)M^4\exp\left(-\frac{Q^2}{\sigma^2}\right)$, $\frac{V''}{V}$ becomes 
\begin{eqnarray}
\frac{V''}{V} = -\frac{2}{\sigma^2}+\frac{4Q^2}{\sigma^4} .
\label{SIGV''/V}
\end{eqnarray}
Because $\frac{2Q}{\sigma^2}$ can be derived when $dw_Q/da$ is observed, $\left(\frac{2Q}{\sigma^2}\right)^2=\frac{4Q^2}{\sigma^4}$ is estimated.  
From Eq. (\ref{d^2w/da^2_A}), the second derivative is given by
\begin{eqnarray}
\frac{d^2w_Q}{da^2} &=& \frac{3}{16\pi}\frac{\Omega_Q}{a^2} \left(1-\frac{\Delta}{2}\right) \left[-4\Delta \left(-\frac{2}{\sigma^2}+\frac{4Q^2}{\sigma^4}\right) \right. \nonumber \\
&+& \left.6\sqrt{\frac{8\pi  \Delta}{3\Omega_Q}} \left \{ (1-\Delta)(6+\Omega_Q)-\frac{1}{3}\Omega_Q\right \}\left(-\frac{2Q}{\sigma^2}\right) \right. \nonumber \\
&+& \left. 4 \left(1-\frac{\Delta}{2}\right)\left(-\frac{2Q}{\sigma^2}\right)^2 + \frac{32\pi \Delta}{\Omega_Q}(7-6\Delta) \right] .
\end{eqnarray}
If $d^2w_Q/da^2$ is observed, $\sigma$ is specified by
\begin{eqnarray}
\sigma^2 &=& \frac{2}{\Delta}\left[\left(1-\frac{\Delta}{2}\right)^{-1}\frac{4\pi a^2}{3\Omega_Q}\frac{d^2 w_Q}{da^2}+\sqrt{\frac{6\pi \Delta}{\Omega_Q}} \left \{ (1-\Delta)(6+\Omega_Q)-\frac{1}{3}\Omega_Q\right \} \left(\frac{2Q}{\sigma^2}\right) \right. \nonumber \\
&-& \left. \left(1-\frac{3\Delta}{2}\right) \left(\frac{2Q}{\sigma^2}\right)^2 - \frac{8\pi}{\Omega_Q}\Delta(7-6\Delta) \right]^{-1} .
\label{sigma2}
\end{eqnarray}
The value $Q$ and $M$ can be also specified by Eq. (\ref{Q/sigma2}) and Eq. (\ref{potential}), respectively.

\subsubsection{Third derivative}
Since the third derivative of the potential is  $V''' = \left(\frac{12Q}{\sigma^4}-\frac{8Q^3}{\sigma^6}\right)M^4\exp\left(-\frac{Q^2}{\sigma^2}\right)$, 
$\frac{V'''}{V}$  is given by
\begin{eqnarray}
\frac{V'''}{V} = \frac{12Q}{\sigma^4}-\frac{8Q^3}{\sigma^6}.
\end{eqnarray}
If the parameters are specified when $dw_Q/da$ and $dw_Q^2/da^2$ are observed, it will become possible to predict the third derivative which is given by 
\begin{eqnarray}
\frac{d^3w_Q}{da^3} &=& A\left[B\left(\frac{12Q}{\sigma^4}-\frac{8Q^3}{\sigma^6}\right)M_{\rm pl}^3 +  C\left(-\frac{2}{\sigma^2}+\frac{4Q^2}{\sigma^4}\right)M_{\rm pl}^2 \right. \nonumber \\
&-& \left . D\frac{2Q}{\sigma^2}M_{\rm pl} + E \left(\frac{4Q}{\sigma^4}-\frac{8Q^3}{\sigma^6}\right)M_{\rm pl}^3 + F\left(-\frac{2Q}{\sigma^2}\right)^2M_{\rm pl}^2 + G \right] .
\end{eqnarray}

 \section{Numerical analysis}
 
  At present the observations of the derivatives of the equation of state are not precise enough to constrain the parameters of the investigated potentials.  However, we present some numerical analysis for the derived equations in this section to infer the observation of $d^3w_Q/da^3$ from 
  the possible range of $dw_Q/da$ and $d^2w_Q/da^2$ in the future. 
  We have not shown the detailed variation of $\Delta$ parameter values and just taken the two typical values $\Delta = 0.1$ and 0.001 at $a=1$.
  For $\Delta = 0.1$, we assume the typical range as $-0.6 < dw_Q/da < 1$ and $-1 < d^2w_Q/da^2 < 1$, and  
  more restrictive ranges are adopted for $\Delta = 0.001$.  
 In the following 3-dimensional presentations, $x, y$ and $z$-axes are taken as $dw_Q/da, d^2w_Q/da^2$ and $d^3w_Q/da^3$, except Fig. 3, 
 where the value of parameter $\alpha$ is taken as z-axis.  The 2-dimensional presentation is made in Fig. 2, where $x$ and $y$-axis are taken as $dw_Q/da$, and $d^2w_Q/da^2$.  
 We have not made numerical analysis for 
 the potential $V= \frac{M^{4+\gamma}}{Q^{\gamma}} \exp\left(\frac{\zeta Q^2}{M_{\rm pl}^2}\right)$, for the predicted values are not derived.   
 The scalar field $Q$ is taken positive for the considered potentials.
 
 \subsection{$V=M^{4+\alpha}/Q^{\alpha}$}

 
 \begin{figure}[ht] 
\includegraphics[clip,width=12cm,height=8cm]{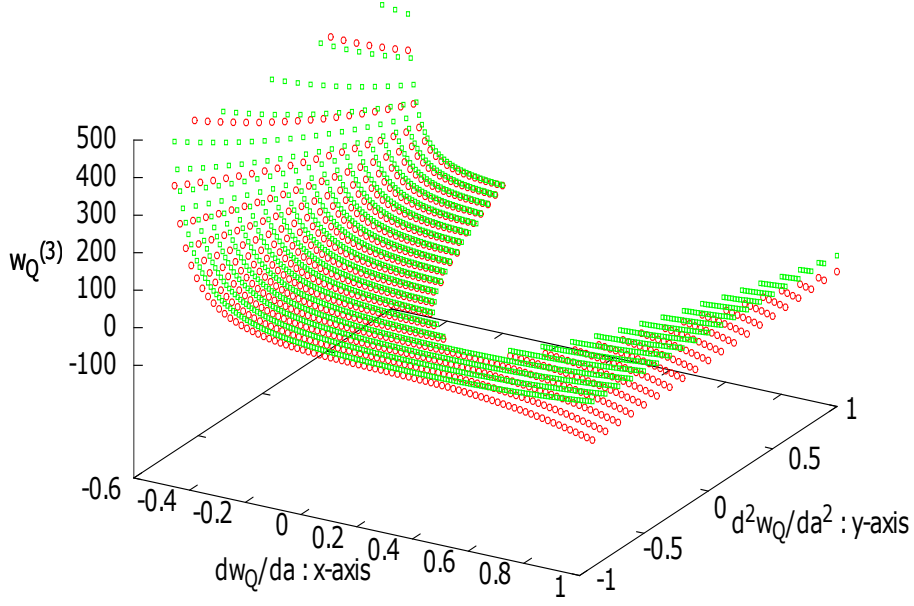}
\caption{The values of $d^3w_Q/da^3=w_Q^{(3)}$ ($z$-axis) are plotted against $dw_Q/da$ ($x$-axis) and $d^2w_Q/da^2$( $y$-axis) for the case of $V=M^{4+\alpha}/Q^{\alpha}$ 
with $\Delta=0.1$ by the green surface (upper side). 
 The case of $V=M^{4}\exp (\beta M/Q)$ with  $\Delta=0.1$ is presented by the red surface (lower side), 
which is described in the following B subsection.  
There are two main forbidden regions where $dw_Q/da < -0.57$ in $x$ coordinate and the parabolic region due to $\alpha < 0$ in $x$-$y$ plane, of which border is shown in Fig. 2 by the red solid curve.
  It should be noted that the same forbidden regions are derived for the potential $V=M^{4}\exp (\beta M/Q)$ which is described later. }
\end{figure}


\begin{figure}[ht]
\includegraphics[clip,width=10cm,height=6cm]{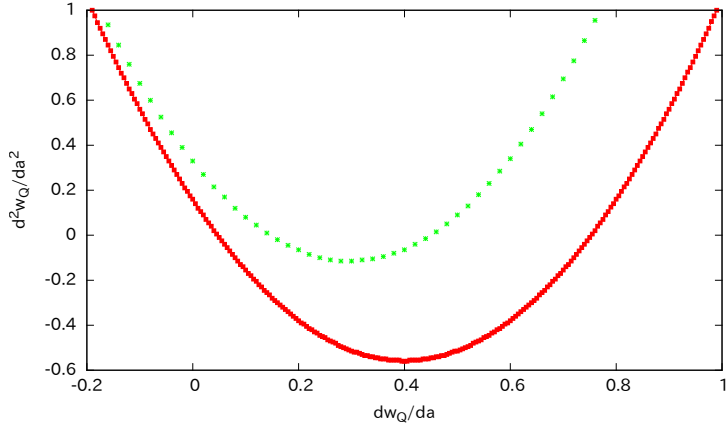}
\caption{The curve for $\alpha =0$ is presented for the case of $V=M^{4+\alpha}/Q^{\alpha}$ with $\Delta=0.1$ by the red solid curve in the $dw_Q/da$ and $d^2w_Q/da^2$ coordinates.
The signature of $\alpha$ will change beyond the parabolic curve.   The green (inner) dotted curve is the case of $V=M^{4}(\cos (Q/f)+1)$ 
with $\Delta=0.1$, which is described in the following C subsection. }   
\end{figure}

 
\begin{figure}[ht]
\includegraphics[clip,width=12cm,height=8cm]{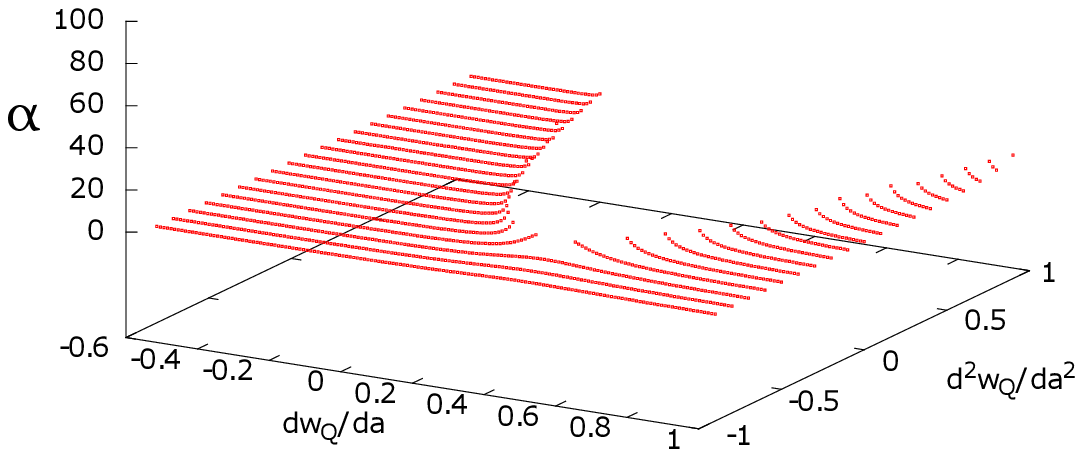}
\caption{The same with Fig. 1 except that the $z$-axis is taken as $\alpha$.  The signature of $\alpha$ will change beyond the boundary ($\alpha = 0)$ of the parabolic curve, shown in Fig. 2 by the red solid curve.
The figure is the case that $\alpha \geq 10$ is eliminated for clarity.  }
\end{figure}

For this inverse power-law potential, the parameter $\alpha$  is positive and $dw_Q/da$ from Eq.(\ref{(6e)}) must satisfy the following equation,
 \begin{eqnarray}
 \frac{dw_Q}{da} > -\frac{6\Delta}{a}(1-\frac{\Delta}{2}).
 \label{dwQ/da}
 \end{eqnarray}
In the following, we notice that this constraint can be applied to all other investigated potentials where all potentials are $V'/V< 0$ in Eq. (\ref{X1}).

For $\Delta$ =0.1 in Fig. 1, this constraint is seen where $dw_Q/da > -0.57$.  
For $d^2w_Q/da^2$, the constraint ($\alpha > $ 0) from Eq. (\ref{4.57a}) causes the depletion of a parabolic inner region in $dw_Q/da$-$d^2w_Q/da^2$ plane of Fig. 1.  
The border is shown in Fig. 2 by the red solid curve.  

 From Eq. (\ref{V2Valpha}), $\alpha$ is given by
  \begin{eqnarray}
 \alpha=\left(\frac{V'}{V}\right)^2\left[\left(\frac{V''}{V}\right)-\left(\frac{V'}{V}\right)^2\right]^{-1}.
 \label{ALPHA}
 \end{eqnarray}
 Using Eqs. (\ref{X1}) and (\ref{Y2}), $\alpha$ is presented by Eq. (\ref{4.57a}).  If we use Eq. (\ref{(6e)}) in Eq. (\ref{4.57a}), $\alpha$ is described by $d^2w_Q/da^2$ and $dw_Q/da$ as
\begin{eqnarray}
\alpha &=& -2\Delta a^2\left(1-\frac{\Delta}{2}\right)\left(a\frac{dw_Q}{da}+6\Delta(1-\frac{\Delta}{2})\right)^{-2}\left[\frac{d^2w_Q}{da^2} \right.   \hspace{3cm} \nonumber \\
&-& \left. \frac{(1-3\Delta/2)}{2\Delta(1-\Delta/2)} \left(\frac{dw_Q}{da}- \frac{3\Delta(1-\Delta/2)}{a(1-3\Delta/2)}(1+\frac{(2-3\Delta)\Omega_Q}{6})\right)^2  \right. \nonumber \\
&+& \left.  \frac{3\Delta(1-\Delta/2)}{8a^2(1-3\Delta/2)} \left(3((1-\Delta)(6+\Omega_Q)-\frac{\Omega_Q}{3})^2-16(7-6\Delta)(1-3\Delta/2) \right)\right]^{-1}.  \label{ALPHAII}
\end{eqnarray}
If we take the square brackets $[ \ \ \ \ ]$ of the above equation as $A \ (=[ \ \ \ \ ])$, the border of the allowed region is described by $A=0$, which is shown in Fig. 2 with the red solid curve.   
It is the parabolic curve in $d^2w_Q/da^2$-$d^2w_Q/da^2$ plane. 
It should be noted that the border is the same for the following exponential type potential $V=M^4\exp(\beta M/Q)$ and the Gaussian type potential $V=M^4\exp(-Q^2/\sigma^2)$.  
For the cosine type potential $V=M^4(\cos(Q/f)+1)$, the similar parabolic border for $\Delta=0.1$ is given in Fig. 2 with green dotted curve which is described in the following C subsection. 

One should note that the allowed and forbidden regions are opposite for the freezing and thawing models. 
At least the forbidden region for $x$-$y$ plane is clear from these figures of potentials $V=M^{4+\alpha}/Q^{\alpha}$ with the green surface (upper side) and $V=M^4\exp(\beta M/Q)$ with the red surface (lower side).  
If the observed values of $dw_Q/da$ and $d^2w_Q/da^2$ are within the forbidden region, 
 the potentials of this type must be excluded.  If the observed values are within the allowed region, the value of $d^3w_Q/da^3$ could be predicted for these potentials.

The value of $\alpha$ is adopted for $z$-axis in Fig. 3, where $x$ and $y$-axes are the same with Fig. 1.  
 It is also seen that the $\alpha $ value increases rapidly when it approaches the parabolic region shown in Fig. 2,
  because the denominator in Eq. (\ref{ALPHAII}) approaches zero and will change the signature beyond the boundary.

For the case of $\Delta=0.001$, almost the same green morphology of Fig. 1 can be seen in Fig. 4, where the scales are different.  
If the observed values are outside of these forbidden regions, the value of $d^3w_Q/da^3$ could be predicted for this potential. 
 The red one is for the case of the potential $V=M^{4}\exp (\beta M/Q)$,  described later. 
 

\begin{figure}[ht]
\includegraphics[clip, width=12cm, height=8cm]{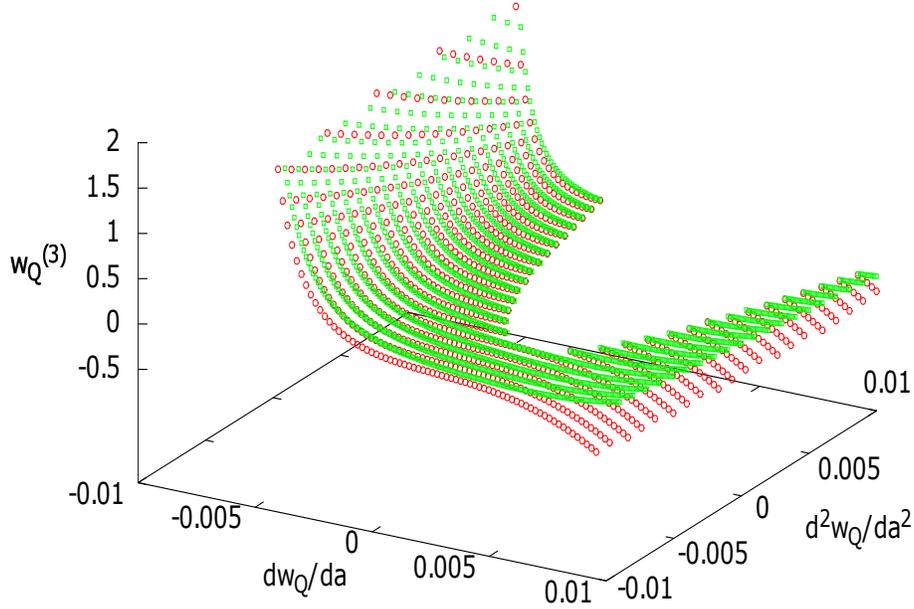}
\caption{The same with Fig. 1 for the case of $\Delta = 0.001$,  where the scale is different.  The values of $d^3w_Q/da^3=w_Q^{(3)}$ ($z$-axis) are plotted against 
$dw_Q/da$ ($x$-axis) and $d^2w_Q/da^2$( $y$-axis).
 There are also two main forbidden regions where $ x < -0.00599$ in $x$ coordinate 
and the parabolic region due to $\alpha < 0$ in $x$-$y$ plane.  The cases for $V=M^{4+\alpha}/Q^{\alpha}$ (up: green surface) and $V=M^4 \exp(\beta M/Q)$ (down: red surface) of $\Delta=0.001$ are shown.  
The differences of the observable values $d^3w_Q/da^3$ can distinguish the potentials. }
\end{figure}

\subsection{$V=M^4 \exp(\beta M/Q)$}


For this exponential type potential, the parameter $ \beta$ is positive and $dw_Q/da$ from Eq. (\ref{FirstDerivativeExp}) must also satisfy Eq. (\ref{dwQ/da}). 
For the case of $\Delta = 0.1$ with the red surface in Fig. 1, this constraint is easily seen where $dw_Q/da >  -0.57.$

By using Eqs. (\ref{V'/V}) and (\ref{V''/V}), $\beta$ is given by
 \begin{eqnarray}
 \beta=\frac{2Q}{M}\left(\frac{V'}{V}\right)^2\left[\left(\frac{V''}{V}\right)-\left(\frac{V'}{V}\right)^2\right]^{-1}.
 \label{BETA}
 \end{eqnarray}
which is similar to Eq. (\ref{ALPHA}).  The criterion of $\beta > 0$ and $\alpha > 0$ is the same with the positive signature of the square brackets [\ \ \ \ ].
The constraint ($Q > 0$ and/or $\beta > 0$) from Eq. (\ref{Mpl/Q}) causes the forbidden region in $dw_Q/da$-$d^2w_Q/da^2$ plane of Fig. 1.  
By comparing Eqs. (\ref{FirstDerivativeExp}) and (\ref{Mpl/Q}) 
with Eqs. (\ref{(6e)}) and (\ref{4.57a}), it is understood that the depletion region is the same for both cases with the constraints $\beta > 0$ and $\alpha > 0$. 
One of the reason is that the first derivative $dw_Q/da$ in Eq. (\ref{X1}) is related to $V'/V$ irrespective of the type of the potential $V$.  
The other reason is that the second derivative $dw^2_Q/da^2$ in Eq. (\ref{d^2w/da^2_A}) and/or (\ref{Y2}) is related to $V'/V$ and $V''/V$ whatever the type of the potential $V$.
The same and a little bit different discussions can be applied to the following potentials, where the forbidden and allowed regions are exchanged with respect to the parabolic border.

For the case of $\Delta=0.001$, almost the same morphology of Fig. 1 with the red surface can be seen in Fig. 4, where the scales are different.  It must be noted $dw_Q/da > -0.00599$.  
 The forbidden region in $x$-$y$ plane is also clear from the figure.  It is also the same that, if the observed values of $dw_Q/da$ and $d^2w_Q/da^2$ are within these regions, 
 the potential of this type must be excluded.  



The differences between the values of $V=M^{4+\alpha}/Q^{\alpha}$ and $V=M^4 \exp (\beta M/Q) $ can be seen in Figs. 1 and 4, 
where $V=M^{4+\alpha}/Q^{\alpha}$  are presented at the upper part (green surface) 
and $V=M^4 \exp(\beta M/Q)$ are shown at the lower part (red surface).  The detailed values of $d^3w_Q/da^3$ have to be presented, if the observational precision of the $dw_Q/da$ and $d^2w_Q/da^2$ could increase.

\subsection{$V=M^4\left[\cos\left(\frac{Q}{f}\right)+1\right]$ }


\begin{figure}[t]
\includegraphics[width=12cm,height=8cm]{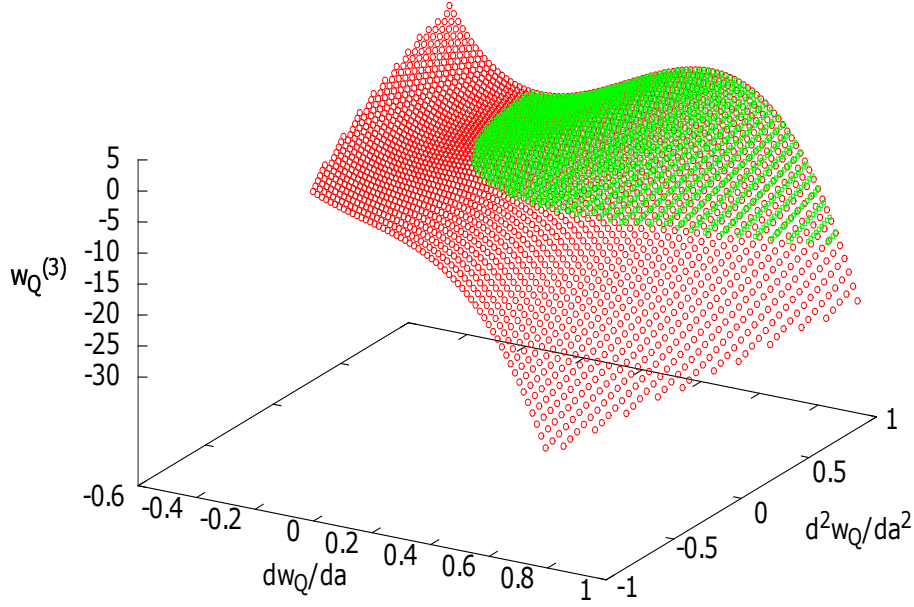}
\caption{The case for $V=M^4\left[\cos\left(\frac{Q}{f}\right)+1\right]$ of $\Delta=0.1$ is presented.  The allowed region is similar to the forbidden parabolic region in Fig. 1.  
The constraint condition $Q > 0$  is similar to the constraint $\alpha > 0$, except the signature.  
The red morphology is the case that, if there is no constraint from the signature of parameters, such as $X, Y >0$ and $X^2-2Y >0$.  
It is shown to understand the allowed green region and its shape. }
\end{figure}

 

 

 \begin{figure}[hb]
\includegraphics[width=17cm,height=8cm]{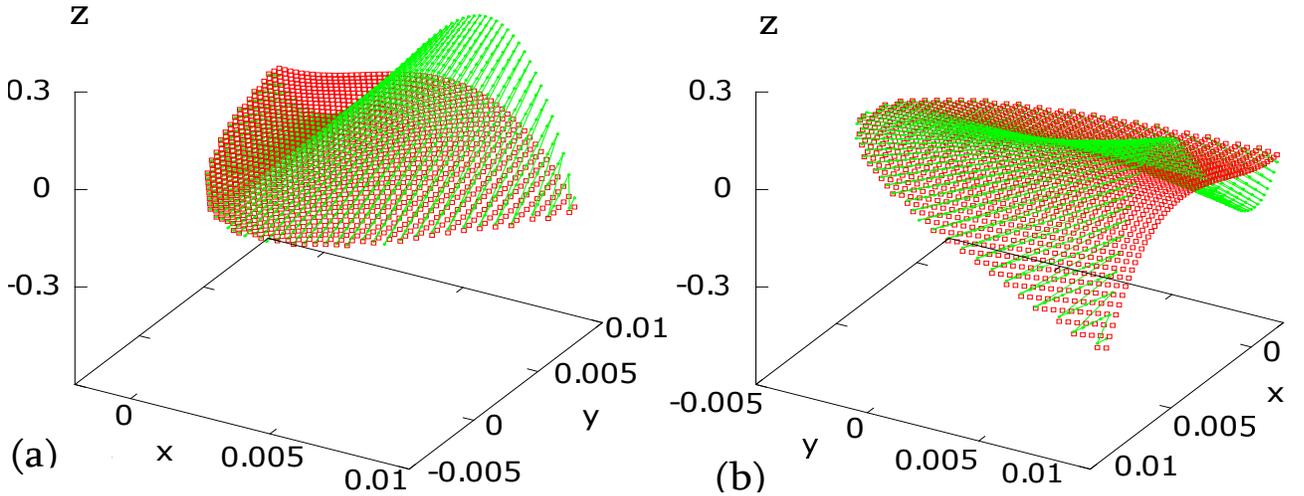}
\caption{The cases for $V=M^4\left[\cos\left(\frac{Q}{f}\right)+1\right]$ (red surface) and $V=M^4\exp \left(-\frac{Q^2}{\sigma ^2} \right)$ (green surface) of $\Delta=0.001$ are shown.  
 The $x, y$ and $z$-axes are taken as $dw_Q/da, d^2w_Q/da^2$ and $d^3w_Q/da^3$, respectively.  The data of Fig. 6 (b) is the same for Fig. 6 (a).  
Fig. 6 (b) is displayed by the clock-wise rotation of Fig. 6 (a) around $z$-axis by 90 degrees.  It can be seen in Fig. 6 (b) that the green surface goes through down the red surface around $x \approx 0$.  
 The differences of the observable values $d^3w_Q/da^3$ can distinguish the potentials.  As $\Delta=0.001$, it must be noted that the differences between the borders estimated from 
 Eqs. (\ref{ALPHAII}) and (\ref{YNEGATIVE}) for both potentials are very small.}
\end{figure}

 For this cosine type potential, the parameters $f$ is positive and $X$ is negative from Eq. (\ref{PN1}).  Then $dw_Q/da$ from Eq. (\ref{COSX}) must also satisfy Eq. (\ref{dwQ/da}).  
 Considering the constraint given by Eq. (\ref{YNEG}) that $Y$ must be negative, Eq. (\ref{COSY}) is arranged using  $d^2w_Q/da^2$ and  $dw_Q/da$ to
 \begin{eqnarray}
Y &=& -\frac{4\pi a^2}{3\Delta \Omega_Q}(1-\frac{\Delta}{2})^{-1}\left[\frac{d^2w_Q}{da^2}-\frac{1}{2\Delta} \left(\frac{dw_Q}{da}-\frac{3\Delta}{a}(1-2\Delta+\frac{(2-3\Delta)\Omega_Q}{6})\right)^2  \right. \nonumber \\
&+& \left.  \frac{3\Delta}{8a^2}\left(3((1-\Delta)(6+\Omega_Q)-\frac{\Omega_Q}{3})^2-16(7-6\Delta)(1-\Delta/2)\right)\right],  \label{YNEGATIVE}
\end{eqnarray}
where the square brackets [ \ \ \ \ ] must be positive, which is different from the freezing type potentials (cf. Eq. (\ref{ALPHAII})).  
The curve of  [ \ \ \ \ ] =0 for $\Delta=0.1$ is presented by a green dotted  parabola in Fig. 2.

The value of $d^3w_Q/da^3$ for $\Delta$ =0.1 is presented in Fig. 5 by the green surface.  
If we remove the constraints of $ X < 0$, $Y < 0$, and $X^2-2Y > 0$ from Eqs. (\ref{PN1}), (\ref{YNEG}), and (\ref{sqrt}), the values of $d^3w_Q/da^3$ have extended to the red one.
The constraint $X < 0$ has not changed the morphology.  Although the constraint $X^2-2Y > 0$ has similar effect, the most severe constraint comes from $Y < 0 $ which has excluded the red part in Fig. 5.  


 For the case of $\Delta=0.001$, almost the same green morphology of Fig. 5 can be seen in Fig. 6 (a) and (b) with the red surface, where the scales are different. 
 The data is the same for Fig. 6 (a) and (b).  Fig. 6 (b) is given by the clock-wise rotation of Fig. 6 (a) around $z$-axis by 90 degrees. 
 It can be seen in Fig. (b) that the green surface sinks below the red surface around $x \approx 0$.  
 The allowed region for $x$-$y$ plane is also clear from these figures.  If the observed values of $dw_Q/da$ and $d^2w_Q/da^2$ are within these regions, 
 the value of $d^3w_Q/da^3$ can be predicted for this potential from the figures.  It should be noted that the differences between the borders estimated from 
 Eqs. (\ref{ALPHAII}) and (\ref{YNEGATIVE}) are very small for $\Delta=0.001$.

\subsection{$V=M^4\exp \left(-\frac{Q^2}{\sigma ^2} \right)$}


\begin{figure}[hc]
\includegraphics[width=12cm,height=8cm]{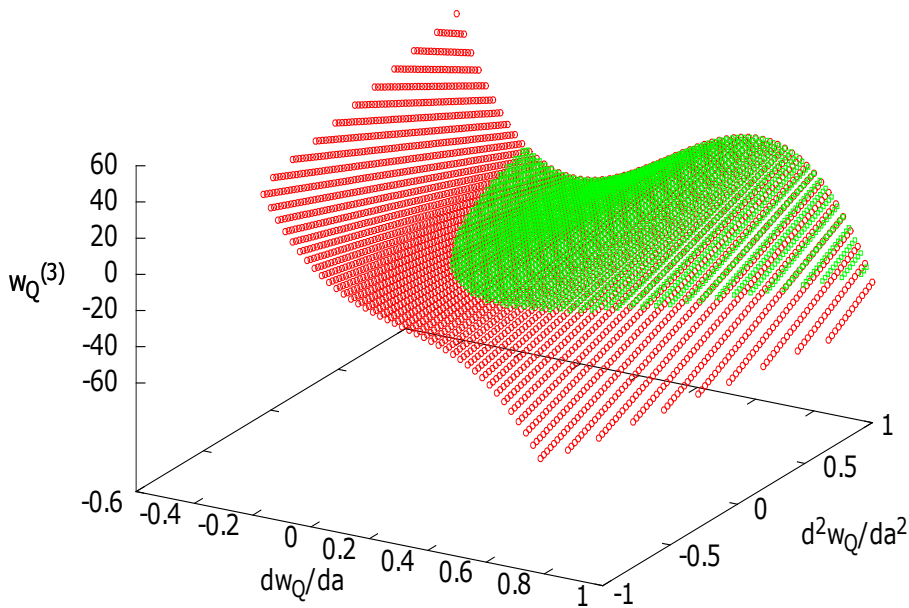}
\caption{The case for $V=M^4\exp \left(-\frac{Q^2}{\sigma ^2} \right)$ of $\Delta=0.1$ is presented. The green surface is the allowed region.  
The red surface is displayed for the understanding of the global values of $d^3w_Q/da^3$. }
\end{figure}

For this Gaussian type potential, there is also the constraint that parameter is positive ( $ \sigma^2  > 0 $).  
Then from Eq. (\ref{Q/sigma2}), $dw_Q/da$ must also satisfy Eq. (\ref{dwQ/da}).  By using Eqs. (\ref{SIGV'/V}) and (\ref{SIGV''/V}), $\sigma^2$ is given by
 \begin{eqnarray}
 \sigma^2=-2\left[\left(\frac{V''}{V}\right)-\left(\frac{V'}{V}\right)^2\right]^{-1}.
 \label{SIGMA}
 \end{eqnarray}
 The constraint that the square brackets [\ \ \ \ ] of the above equation must be negative is the opposite signature for the case of Eqs. (\ref{ALPHA}) and (\ref{BETA}).  
 Comparing Eqs. (\ref{sigma2}) and (\ref{Q/sigma2}) with Eqs. (\ref{4.57a}) and (\ref{(6e)}), it is understood that the border is the same. 
 The border is shown in the parabolic red solid curve in Fig. 2 for the case of $\Delta=0.1$, however the allowed and forbidden regions are opposite.  
 
 



For $\Delta =0.1 $ in Fig. 7, this constraint is satisfied for the green part.  
Even if the constraints of $ 2Q/\sigma ^2 > 0 $ from Eq. (\ref{Q/sigma2}) is removed, the figure does not change.  
However, if we remove the constraint of $\sigma^2 > 0$ 
from  Eq. (\ref{sigma2}), the figure has extended to the red region.

 For the case of $\Delta=0.001$, almost the same morphology of Fig. 7 can be seen in Fig.6 with the green one, where the scales are different. 
 The allowed region for $x$-$y$ plane is also clear from the figure, which also means that the square brackets in Eq. (\ref{sigma2}) must be positive.  
 If the observed values of $dw_Q/da$ and $d^2w_Q/da^2$ are within these regions, 
 the value of $d^3w_Q/da^3$ could be predicted from the figures for this potential.

The differences between the values of $V=M^4\left[\cos\left(\frac{Q}{f}\right)+1\right]$ and 
$V=M^4\exp \left(-\frac{Q^2}{\sigma ^2} \right)$  for the case of $\Delta=0.001$ can be seen in Figs. 6 (a) and (b), where those of $V=M^4\left[\cos\left(\frac{Q}{f}\right)+1\right]$ are presented with the red surface.

\section{Conclusions and discussion}
 If more details of the accelerating universe are observed, it will be important to find out the time variation of the equation of state to understand the so-called dark energy.  
 Various potentials are proposed to explain the acceleration in the context of quintessence with a single scalar field.  
 It is necessary to distinguish which type of the potential will be the theory to explain the expansion \cite{10z,10a}.  
 To differentiate the potential, it is necessary to specify the parameters of the potential.  
 In this paper we have studied the method to find out the potential and calculated 
 the third derivative of the equation of state for several potentials for this purpose.

At present, backward observation of large-scale structure of the universe has been undertaken to estimate $w_Q$ at the age $(1+z)$ \cite{10}.
Actually, the values of $w(a=a_0)$, $dw/da$ \cite{7,8,9} and  $d^2w/da^2$ \cite{6} have been pursued to be determined from the observation:
\begin{align}
w(a)=w(a=a_0)+\frac{dw}{da}da+\frac{1}{2}\frac{d^2w}{da^2}(da)^2 .
\end{align}

Since observations of the third and higher derivatives of $w$ can be expected in the future
\begin{align}
w(a)=w(a_0=1)+\frac{dw}{da}da+\frac{1}{2}\frac{d^2w}{da^2}(da)^2+\frac{1}{3!}\frac{d^3w}{da^3}(da)^3+\cdots  ,
\end{align}
we have calculated the third derivative of $w$ for general potential $V$ and applied to five typical potentials. 
Three are the freezing model; the inverse power type ($V=M^{4+\alpha}/Q^{\alpha}$),  the exponential type 
($V=M^4\exp{(\beta M/Q)}$), and the mixed type ($V=M^4/Q^{\gamma}\exp(\zeta Q^2/ M_{\rm pl}^2)$, 
and two are the thawing model; 
the PNGB type 
$(V=M^{4}(\cos(Q/f)+1))$, 
and the Gaussian type ($V=M^4\exp{(Q^2/\sigma^2)}$).

 Numerical analysis for $d^3w/da^3$ are made under some specified parameters in the investigated potentials, except the mixed one. 
 For each potential, there is a forbidden region in the expected values of $dw/da$ and $d^2w/da^2$, which are presented in Sect. V.

There is common forbidden region that $dw/da > -6\Delta(1-\Delta/2)$, which is due to the same signature of $V'/V< 0$ in Eq.(\ref{X1}) for the adopted potentials.
Another interesting point is that the forbidden regions for the freezing type potentials are allowed region for the thawing type potentials and the reverse is also true. 
It is possible to distinguish the potentials among each type due to the different predicted values of $d^3w/da^3$, 
however it is necessary to make accurate observations for the values of $dw/da$, $d^2w/da^2$, $d^3w/da^3$ and other parameters such as $\Delta$.

 About the adopted potentials, the four of them have two parameters and one has three parameters to identify the form.  
 As stated before, if there is  $n$ parameters of the potential to specify, it is necessary to observe the $n$ derivatives of $w_Q$ ($dw_Q/da, d^2w_Q/da^2, \cdots, d^nw_Q/da^n$). 
If the potential is specified, it becomes possible to predict the $n+1$ derivative of $w_Q$ ($d^{n+1}w_Q/da^{n+1}$).

 For example about the inverse power-law potential ($V=M^{4+\alpha}/Q^{\alpha}$), 
the observed first and second derivatives ($dw_Q/da$, $d^2w_Q/da^2$) with $H, w_Q,$ and $\Omega_Q$ can determine the two parameters of the potential $M$
 and $\alpha$.  The point is that the third derivative ($d^3w_Q/da^3$) is described by the current values of parameters  $\Omega_Q $, $w_Q $, 
 and its time derivatives, $dw_Q/da$, and $d^2w_Q/da^2$.  The values of $d^3w/da^3$ are calculated under some specified parameters in Figs. 1, 3 and 6.
 If it is predicted value, it could be understood that the dark energy would be described by the quintessence 
 with a single scalar field of this potential.  At least it will satisfy the necessary condition.  
 It seems to be difficult to define the sufficient condition for the model of the dark energy.  
 However, in principle, the higher derivatives $d^nw_Q/da^n$ ($n \geq 3$) can be predicted from the specific potentials.　


 The evolution of forward and/or backward time variation can be analyzed at some fixed time point. 
If the potential is known, the evolution will be estimated from values $Q$ 
and $\dot {Q}$ at this point, because the equation for scalar field is the second derivative equation as in Eq. (\ref{Qfield}).

 About the derivative of $w_Q$, if $w_Q(z_i)$ at redshit $z_i$ is observed, the derivative of $w_Q$ is given by 
 $dw_Q/da \simeq (w_Q(z_i)-w_Q(z_{i+1}))/(a(z_i)-a(z_{i+1}))$, for $a(z_i) > a(z_{i+1})$. 
 The $n$-th derivative of $w_Q$ can be derived through the observation $w_Q(z_i)$, where $i$ takes  $ 1, 2, \cdots, n+1 $, respectively.  
 The differences of $w_Q(z_i)$ can give the higher derivative of $w_Q(z_i)$.
If we get the form of the potential, we could predict any higher derivative of $w_Q$ by the observables through the method developed in the paper \cite {6} and this paper.　

If $\Delta < 0$, we must consider fully different models such as phantoms \cite{19}, quintom \cite{20} or k-essence \cite{21}. 
There are other models which are proposed to explain dark energy such as chameleon field \cite{22}, tachyon field \cite{23}, dilaton field \cite{24},
 holographic dark energy \cite{25},
modified gravity theory \cite{26}, and so on \cite{27}.
These models should be considered to parameterize the characteristic features in relation with the high derivatives of the accelerated expansion velocity and observable quantities.

\vspace{1cm}
\hspace{-0.5cm} {\large{\bf Appendix}}

\appendix

\section{Derivation of the third derivative of $w_Q$ in Eq. (\ref{d3w/da3})}
Here, we describe the calculation to derive the third derivative of $w_Q$.  It should be noted that the third derivative is denoted by the superscript by (3) and 
the differences should be noted by parentheses ( ), curly brackets \{  \} , and brackets [ ].

From Eq. (\ref{d^2w/da^2}),  the third derivative  is given as
\begin{eqnarray}
\frac{d^3w_Q}{da^3} &=& \frac{dt}{da}\frac{d}{dt}\left[\frac{(\ddot{p_Q}\rho_Q - p_Q \ddot{\rho_Q})\dot{a}\rho_Q^2 - (\dot{p_Q}\rho_Q - p_Q \dot{\rho_Q})(\ddot{a}\rho_Q^2 + 2\dot{a}\rho_Q\dot{\rho_Q})}{\dot{a}^3\rho_Q^4}\right] \nonumber \\
&=& \frac{1}{\dot{a}^7 \rho_Q^8}\left[\{ (\ddot{p_Q}\rho_Q - p_Q \ddot{\rho_Q})\dot{a}\rho_Q^2 - (\dot{p_Q}\rho_Q - p_Q \dot{\rho_Q})(\ddot{a}\rho_Q^2 + 2\dot{a}\rho_Q\dot{\rho_Q}) \} ^{\cdotp} \ \dot{a}^3\rho_Q^4 \right. \nonumber \\
&-& \left. \{(\ddot{p_Q}\rho_Q - p_Q \ddot{\rho_Q})\dot{a}\rho_Q^2 - (\dot{p_Q}\rho_Q - p_Q \dot{\rho_Q})(\ddot{a}\rho_Q^2 + 2\dot{a}\rho_Q\dot{\rho_Q}) \}(\dot{a}^3\rho_Q^4)^{\cdotp} \right] , \nonumber \\
\end{eqnarray}
where the dot symbol $  ^{\cdotp} $ means the derivative with time.

The derivative of the first term in the above equation becomes  
\begin{eqnarray*}
\{ (\ddot{p_Q}\rho_Q - p_Q \ddot{\rho_Q})\dot{a}\rho_Q^2 - (\dot{p_Q}\rho_Q - p_Q \dot{\rho_Q})(\ddot{a}\rho_Q^2 + 2\dot{a}\rho_Q\dot{\rho_Q}) \} ^{\cdotp}
\end{eqnarray*}
\begin{eqnarray}
&=& (\ddot{p_Q}\rho_Q - p_Q \ddot{\rho_Q})^{\cdotp}\dot{a}\rho_Q^2 + (\ddot{p_Q}\rho_Q - p_Q \ddot{\rho_Q})(\dot{a}\rho_Q^2) ^{\cdotp} \nonumber \\
&-& (\dot{p_Q}\rho_Q - p_Q \dot{\rho_Q})^{\cdotp}(\ddot{a}\rho_Q^2 + 2\dot{a}\rho_Q\dot{\rho_Q})-(\dot{p_Q}\rho_Q - p_Q \dot{\rho_Q})(\ddot{a}\rho_Q^2 + 2\dot{a}\rho_Q\dot{\rho_Q}) ^{\cdotp} \nonumber \\
&=& (p_Q^{(3)}\rho_Q+\ddot{p_Q}\dot{\rho_Q}-\dot{p_Q} \ddot{\rho_Q}-p_Q\rho_Q^{(3)})\dot{a}\rho_Q^2-(\dot{p_Q}\rho_Q 
- p_Q\dot{\rho_Q})(a^{(3)}\rho_Q^2+4\ddot{a}\rho_Q\dot{\rho_Q}+2\dot{a}\dot{\rho_Q}^2+2\dot{a}\rho_Q\ddot{\rho_Q} ) , \nonumber \\
\label{ThirdDeriFirst}
\end{eqnarray}
then
\begin{eqnarray}
\frac{d^3w_Q}{da^3} 
&=& \frac{1}{(\dot{a}\rho_Q)^5}\left[ \left \{ (p_Q^{(3)}\rho_Q+\ddot{p_Q}\dot{\rho_Q}-\dot{p_Q} \ddot{\rho_Q}-p_Q\rho_Q^{(3)})\dot{a}\rho_Q^2 \right. \right. \nonumber \\
&-& \left. \left. (\dot{p_Q}\rho_Q - p_Q\dot{\rho_Q})(a^{(3)}\rho_Q^2+4\ddot{a}\rho_Q\dot{\rho_Q}+2\dot{a}\dot{\rho_Q}^2+2\dot{a}\rho_Q\ddot{\rho_Q} ) \right \}\dot{a}\rho_Q \right. \nonumber \\
&-& \left. \{ (\ddot{p_Q}\rho_Q - p_Q \ddot{\rho_Q})\dot{a}\rho_Q^2 - (\dot{p_Q}\rho_Q - p_Q \dot{\rho_Q})(\ddot{a}\rho_Q^2 + 2\dot{a}\rho_Q\dot{\rho_Q})\}(3\ddot{a}\rho_Q+4\dot{a}\dot{\rho_Q}) \right] .\nonumber \\
\label{d3w/da3Appe}
\end{eqnarray}
Since the second derivatives are described in the paper \cite{6},  the third derivatives of $p_Q,$ and $\rho_Q$ are shown here as, 
\begin{eqnarray}
p_Q^{(3)} &=& \left(-3\frac{a^{(3)}a-\ddot{a}\dot{a}}{a^2}+42H\dot{H}-2\dot{Q}V''' \right)\dot{Q}^2 + \left(-3\frac{\ddot{a}}{a}+21H^2-2V''\right)2\dot{Q}(-3H\dot{Q}-V') \nonumber \\
&+& 12(\dot{H}\dot{Q}V'+H\ddot{Q}V'+H\dot{Q}^2V'')+4V'\dot{Q}V'' \nonumber \\
&=& \left(-3\frac{a^{(3)}}{a}+63H\frac{\ddot{a}}{a}-168H^3+24HV''-2\dot{Q}V''' \right)\dot{Q}^2 \nonumber \\
&+& \left(18\frac{\ddot{a}}{a}-90H^2 +8V''\right)V'\dot{Q}-12HV'^2 .
\end{eqnarray}
\begin{eqnarray}
\hspace{-2.0cm}\rho_Q^{(3)} &=& \left(-3\frac{a^{(3)}a-\ddot{a}\dot{a}}{a^2}+42H\dot{H}\right)\dot{Q}^2+\left(-3\frac{\ddot{a}}{a}+21H^2\right)2\dot{Q}(-3H\dot{Q}-V') \nonumber \\
&+& 6(\dot{H}\dot{Q}V'+H\ddot{Q}V'+H\dot{Q}^2V'') \nonumber \\
&=& \left(-3\frac{a^{(3)}}{a}+63H\frac{\ddot{a}}{a}-168H^3+6HV''\right)\dot{Q}^2+ \left(12\frac{\ddot{a}}{a}-66H^2\right)V'\dot{Q}-6HV'^2 .
\end{eqnarray}
Hereafter we calculate  each term of  Eq. (\ref{d3w/da3Appe}) consequently.  Beforehand we show the necessary peaces as \cite{6}
\begin{eqnarray*}
\frac{\ddot{a}}{a} &=& 4\pi G \rho_c\left[\Omega_Q\left(\frac{2}{3}-\Delta\right)\right], \\
\dot{Q}^2 &=& \rho_c \Omega_Q\Delta = \rho_Q \Delta , \\
\dot{p_Q} &=& -3H\rho_Q \Delta-2V'\dot{Q} , \\
\dot{\rho_Q} &=& -3H\rho_Q\Delta , \\
\rho_c &=& \frac{3H^2}{8\pi G}\rightarrow H^2 = \frac{8\pi G}{3}\rho_c .
\end{eqnarray*}
Next we calculate the elements appeared in this section as
\begin{eqnarray}
\frac{a^{(3)}}{a} &=& \frac{d}{dt}\left(\frac{\ddot{a}}{a}\right)+\frac{\ddot{a}}{a}\frac{\dot{a}}{a} \nonumber \\
&=& 4\pi G\left(-\frac{\dot{\rho_Q}}{3}-\dot{p_Q}\right)+4\pi GH\rho_c \left[\Omega_Q\left(\frac{2}{3}-\Delta \right)\right] \nonumber \\
&=& 4\pi G\left \{4H\rho_Q\Delta+2V'\sqrt{\rho_Q\Delta}+H\rho_c \left[\Omega_Q\left(\frac{2}{3}-\Delta \right)\right]\right \} ,
\end{eqnarray}
\begin{eqnarray}
\ddot{p_Q} = \left \{-12\pi G\rho_c\left[\Omega_Q\left(\frac{2}{3}-\Delta \right)\right]+21H^2-2V'' \right \} \rho_Q \Delta+12HV'\sqrt{\rho_Q\Delta}+2V'^2 , \nonumber \\
\end{eqnarray}
\begin{eqnarray}
\ddot{\rho_Q} = \left \{-12\pi G\rho_c\left[\Omega_Q\left(\frac{2}{3}-\Delta \right)\right]+21H^2 \right \} \rho_Q \Delta+6HV'\sqrt{\rho_Q\Delta} .
\end{eqnarray}

Hereafter we calculate each term in the bracket [ \ ] of Eq. (\ref{d3w/da3Appe}) separately. 

At first, the second term becomes  
\begin{eqnarray*}
\{(\ddot{p_Q}\rho_Q - p_Q \ddot{\rho_Q})\dot{a}\rho_Q^2 - (\dot{p_Q}\rho_Q - p_Q \dot{\rho_Q})(\ddot{a}\rho_Q^2 + 2\dot{a}\rho_Q\dot{\rho_Q})\}(3\ddot{a}\rho_Q+4\dot{a}\dot{\rho_Q}) . \\
\end{eqnarray*}
Since the left curly bracket \{ \  \} part is calculated in the paper \cite{6}, the right parenthesis ( \ \ \ ) part is calculated as
\begin{eqnarray}
3\ddot{a}\rho_Q+4\dot{a}\dot{\rho_Q} &=& a\left(3\frac{\ddot{a}}{a}\rho_Q+4\frac{\dot{a}}{a}\dot{\rho_Q}\right) \nonumber \\ 
&=& 12\pi G \rho_c\rho_Qa\left[\Omega_Q\left(\frac{2}{3}-\Delta \right)-\frac{8}{3}\Delta\right] .
\end{eqnarray}
Then the second term within square bracket [ \ \ \  ] of Eq. (\ref{d3w/da3Appe}) becomes 
\begin{eqnarray}
& \ & \hspace{-0.5cm}12\pi G\rho_c\rho_Q a\left[\Omega_Q\left(\frac{2}{3}-\Delta \right)-\frac{8}{3}\Delta\right] \times \frac{1}{2}a\rho_Q^5\Delta^2\left(1-\frac{\Delta}{2}\right)\sqrt{\frac{8\pi G}{3}\rho_c}\left[-\frac{4}{\Delta} \frac{V''}{V} \right. \nonumber \\
&+& \left. 6\sqrt{\frac{8\pi G}{3\Omega_Q}}\frac{1}{\Delta^{\frac{3}{2}}}\left \{ (1-\Delta)(6+\Omega_Q)-\frac{1}{3}\Omega_Q\right \} \frac{V'}{V}+\left(\frac{2}{\Delta}\right)^2 \left(1-\frac{\Delta}{2}\right)\left(\frac{V'}{V}\right)^2 + \frac{32\pi G}{\Delta \Omega_Q}(7-6\Delta) \right] . \nonumber \\
\label{ThirdDeriSec}
\end{eqnarray}
Next we consider the first term; 
\begin{eqnarray*}
\{(p_Q^{(3)}\rho_Q+\ddot{p_Q}\dot{\rho_Q}-\dot{p_Q} \ddot{\rho_Q}-p_Q\rho_Q^{(3)})\dot{a}\rho_Q^2-(\dot{p_Q}\rho_Q - p_Q\dot{\rho_Q})(a^{(3)}\rho_Q^2+4\ddot{a}\rho_Q\dot{\rho_Q}+2\dot{a}\dot{\rho_Q}^2+2\dot{a}\rho_Q\ddot{\rho_Q} )\} \dot{a}\rho_Q .
\end{eqnarray*}
Two elements of the above equation are following,
\begin{eqnarray*}
\dot{a}\rho_Q^2 &=& a\sqrt{\frac{8\pi G}{3}\rho_c}\rho_Q^2  ,\\
\dot{p_Q}\rho_Q - p_Q \dot{\rho_Q} &=& -2V\rho_Q \sqrt{\rho_c}\left(\sqrt{24\pi G}\Delta + \frac{V'}{V}\sqrt{\Delta \Omega_Q} \right) , 
\end{eqnarray*}
then we need to calculate the following two parts 
\[
   \begin{array}{ll}
   {\rm (i) \ \ }    p_Q^{(3)}\rho_Q+\ddot{p_Q}\dot{\rho_Q}-\dot{p_Q} \ddot{\rho_Q}-p_Q\rho_Q^{(3)} , \\
   {\rm (ii) \ \ }    a^{(3)}\rho_Q^2+4\ddot{a}\rho_Q\dot{\rho_Q}+2\dot{a}\dot{\rho_Q}^2+2\dot{a}\rho_Q\ddot{\rho_Q} .\\
  \end{array}
\]
The second part (ii) becomes 
\begin{eqnarray*}
\hspace{-0.8cm} a^{(3)}\rho_Q^2+4\ddot{a}\rho_Q\dot{\rho_Q}+2\dot{a}\dot{\rho_Q}^2+2\dot{a}\rho_Q\ddot{\rho_Q} &=& a\left(\frac{a^{(3)}}{a}\rho_Q^2+4\frac{\ddot{a}}{a}\rho_Q\dot{\rho_Q}+2\frac{\dot{a}}{a}\dot{\rho_Q}^2+2\frac{\dot{a}}{a}\rho_Q\ddot{\rho_Q}\right)
\end{eqnarray*}
\begin{eqnarray}
&=& a\left[4\pi G\left \{4H\rho_Q\Delta+2V'\sqrt{\rho_Q\Delta}+H\rho_c \left[\Omega_Q\left(\frac{2}{3}-\Delta \right)\right]\right \}\rho_Q^2 \right. \nonumber \\
&+& \left. 16\pi G \rho_c\left[\Omega_Q\left(\frac{2}{3}-\Delta\right)\right]\rho_Q(-3H\rho_Q\Delta) + 2H(-3H\rho_Q\Delta)^2 \right. \nonumber \\
&+& \left. 2H\rho_Q\left(\left \{-12\pi G\rho_c\left[\Omega_Q\left(\frac{2}{3}-\Delta \right)\right]+21H^2 \right \} \rho_Q \Delta+6HV'\sqrt{\rho_Q\Delta}\right) \right] \nonumber \\
\nonumber \\
&=& a\left[\Omega_Q\left(\frac{2}{3}-\Delta\right)\left(4\pi GH\rho_Q^2\rho_c-72\pi G\rho_c\rho_Q^2H\Delta\right) \right. \nonumber \\
&+& \left. 16\pi G \rho_Q^3H\Delta+8\pi G\rho_Q^2V'\sqrt{\rho_Q\Delta}+18H^3\rho_Q^2\Delta^2+42H^3\rho_Q^2\Delta+12H^2\rho_QV'\sqrt{\rho_Q\Delta}\right] \nonumber \\
\nonumber \\
&=& 4\pi G\rho_c\rho_Q^2a\left[H\Omega_Q\left(\frac{2}{3}-\Delta\right)(1-18\Delta)+4H\Delta(\Omega_Q+3\Delta+7)+2\frac{V'}{V}\left(1-\frac{\Delta}{2}\right)\sqrt{\Delta\rho_Q}(\Omega_Q+4)\right]  .\nonumber \\
\end{eqnarray}
The right hand part of the first term becomes 
\begin{eqnarray*}
\hspace{-4cm}(\dot{p_Q}\rho_Q - p_Q\dot{\rho_Q})(a^{(3)}\rho_Q^2+4\ddot{a}\rho_Q\dot{\rho_Q}+2\dot{a}\dot{\rho_Q}^2+2\dot{a}\rho_Q\ddot{\rho_Q})
\end{eqnarray*}
\begin{eqnarray}
&=& -\frac{8\pi Ga \rho_Q^5 \rho_c^{\frac{1}{2}}}{\Omega_Q}\left(1-\frac{\Delta}{2}\right) \left[\sqrt{24\pi G}\Delta + \frac{V'}{V}\sqrt{\Delta \Omega_Q} \right] \left[H\Omega_Q\left(\frac{2}{3}-\Delta\right)(1-18\Delta) \right. \nonumber \\
&+& \left. 4H\Delta(\Omega_Q+3\Delta+7)+2\frac{V'}{V}\left(1-\frac{\Delta}{2}\right)\sqrt{\Delta\rho_Q}(\Omega_Q+4)\right] .
\end{eqnarray}
At last we calculate the first term as
\begin{eqnarray*}
\hspace{-4cm} p_Q^{(3)}\rho_Q+\ddot{p_Q}\dot{\rho_Q}-\dot{p_Q} \ddot{\rho_Q}-p_Q\rho_Q^{(3)}
\end{eqnarray*}
\begin{eqnarray}
&=& \left[ \left(-3\frac{a^{(3)}}{a}+63H\frac{\ddot{a}}{a}-168H^3+24HV''-2\dot{Q}V''' \right)\dot{Q}^2 \right. \nonumber \\
&+& \left. \left(18\frac{\ddot{a}}{a}-90H^2 +8V''\right)V'\dot{Q}-12HV'^2\right] \left(\frac{1}{2}\dot{Q}^2+V\right) \nonumber \\
&+& \left[ \left(-3\frac{\ddot{a}}{a}+21H^2-2V''\right)\dot{Q}^2 + 12H\dot{Q}V' +2V'^2\right](-3H\dot{Q}^2) \nonumber \\
&-& (-3H\dot{Q}^2-2V'\dot{Q})\left[ \left(-3\frac{\ddot{a}}{a} +21H^2\right)\dot{Q}^2+6H\dot{Q}V'\right] \nonumber \\
&-& \left(\frac{1}{2}\dot{Q}^2-V\right)\left[\left(-3\frac{a^{(3)}}{a}+63H\frac{\ddot{a}}{a}-168H^3+6HV''\right)\dot{Q}^2 \right. \nonumber \\
&+& \left. \left(12\frac{\ddot{a}}{a}-66H^2\right)V'\dot{Q}-6HV'^2\right] .
\end{eqnarray}
We concentrate the first and fourth terms and sort out as
\begin{eqnarray*}
p_Q^{(3)}\rho_Q-p_Q\rho_Q^{(3)} &=& \left[(18HV''-2\dot{Q}V''')\dot{Q}^2+\left(6\frac{\ddot{a}}{a}V'-24H^2V'+8V'V''\right)\dot{Q}-6HV'^2\right] \frac{1}{2}\dot{Q}^2 \nonumber \\
&+& \left[\left(-6\frac{a^{(3)}}{a}+126H\frac{\ddot{a}}{a}-336H^3+30HV''-2\dot{Q}V'''\right)\dot{Q}^2 \right. \nonumber \\
&+& \left. \left(30\frac{\ddot{a}}{a}V'-156H^2V'+8V'V''\right)\dot{Q}-18HV'^2\right]V \nonumber \\
\end{eqnarray*}
\vspace{-1.0cm}
\begin{eqnarray}
&=& \left[18\frac{H\dot{Q}^2}{V}\frac{V''}{V}-2\frac{\dot{Q}^3}{V}\frac{V'''}{V}+6\frac{\ddot{a}}{a}\frac{\dot{Q}}{V}\frac{V'}{V}-24\frac{H^2\dot{Q}}{V}\frac{V'}{V}+8\dot{Q}\frac{V'}{V}\frac{V''}{V}-6H\left(\frac{V'}{V}\right)^2\right] \frac{1}{2}\dot{Q}^2V^2 \nonumber \\
&+& \left[-6\frac{a^{(3)}}{a}\frac{\dot{Q}^2}{V^2}+126H\frac{\ddot{a}}{a}\frac{\dot{Q}^2}{V^2}-336\frac{H^3\dot{Q}^2}{V^2}+30\frac{H\dot{Q}^2}{V}\frac{V''}{V}-2\frac{\dot{Q}^3}{V}\frac{V'''}{V} \right. \nonumber \\
&+& \left. 30\frac{\ddot{a}}{a}\frac{\dot{Q}}{V}\frac{V'}{V}-156\frac{H^2\dot{Q}}{V}\frac{V'}{V}+8\dot{Q}\frac{V'}{V}\frac{V''}{V}-18H\left(\frac{V'}{V}\right)^2\right]V^3 .
\end{eqnarray}
Here we show the necessary parts as 
\begin{eqnarray}
\frac{H\dot{Q}^2}{V} &=& \frac{\rho_Q\Delta \sqrt{\frac{8\pi G}{3}\rho_c}}{\rho_c\Omega_Q\left(1-\frac{\Delta}{2}\right)} = \frac{\Delta \sqrt{\frac{8\pi G}{3}\rho_c}}{1-\frac{\Delta}{2}} ,\\
\frac{\dot{Q}^3}{V} &=& \frac{\rho_Q\Delta \sqrt{\rho_Q\Delta}}{\rho_c\Omega_Q\left(1-\frac{\Delta}{2}\right)} = \frac{\Delta \sqrt{\rho_Q\Delta}}{1-\frac{\Delta}{2}} ,\\
\frac{\ddot{a}}{a}\frac{\dot{Q}}{V} &=& 4\pi G \rho_c\Omega_Q\left(\frac{2}{3}-\Delta \right) \frac{\sqrt{\rho_Q\Delta}}{\rho_c\Omega_Q\left(1-\frac{\Delta}{2}\right)} 
=  \frac{4\pi G \left(\frac{2}{3}-\Delta \right)\sqrt{\rho_Q\Delta}}{1-\frac{\Delta}{2}} , \\
\frac{H^2\dot{Q}}{V} &=& \frac{\frac{8\pi G}{3}\rho_c \sqrt{\rho_Q\Delta}}{\rho_c\Omega_Q\left(1-\frac{\Delta}{2}\right)} = \frac{8\pi G}{3} \frac{\sqrt{\rho_Q\Delta}}{\Omega_Q\left(1-\frac{\Delta}{2}\right)}  ,\\
\frac{a^{(3)}}{a}\frac{\dot{Q}^2}{V^2} &=& 4\pi G\left \{4H\rho_Q\Delta+2V'\sqrt{\rho_Q\Delta}+H\rho_c \left[\Omega_Q\left(\frac{2}{3}-\Delta \right)\right]\right \}\frac{\rho_Q \Delta}{\rho_c^2\Omega_Q^2\left(1-\frac{\Delta}{2}\right)^2}  \nonumber \\
&=& \frac{16\pi G\sqrt{\frac{8\pi G}{3}\rho_c}\Delta^2}{\left(1-\frac{\Delta}{2}\right)^2}+\frac{8\pi G\Delta\sqrt{\rho_Q\Delta}}{1-\frac{\Delta}{2}}\frac{V'}{V}+\frac{4\pi G\Delta \sqrt{\frac{8\pi G}{3}\rho_c}\left(\frac{2}{3}-\Delta \right)}{\left(1-\frac{\Delta}{2}\right)^2}  ,\nonumber 
\end{eqnarray}
\begin{eqnarray}
\frac{\ddot{a}}{a}\frac{H\dot{Q}^2}{V^2} &=& 4\pi G \rho_c\Omega_Q\left(\frac{2}{3}-\Delta \right)\frac{\sqrt{\frac{8\pi G}{3}\rho_c}\rho_Q\Delta}{\rho_c^2\Omega_Q^2\left(1-\frac{\Delta}{2}\right)^2}  \nonumber \\
&=& \frac{4\pi G\left(\frac{2}{3}-\Delta \right)\sqrt{\frac{8\pi G}{3}\rho_c}\Delta}{\left(1-\frac{\Delta}{2}\right)^2} ,  \\
\frac{H^3\dot{Q}^2}{V^2} &=& \frac{\frac{8\pi G}{3}\rho_c\sqrt{\frac{8\pi G}{3}\rho_c} \rho_Q\Delta}{\rho_c^2\Omega_Q^2\left(1-\frac{\Delta}{2}\right)^2} = \frac{8\pi G}{3}\sqrt{\frac{8\pi G}{3}\rho_c}\frac{\Delta}{\Omega_Q\left(1-\frac{\Delta}{2}\right)^2} .
\end{eqnarray}
Then $p_Q^{(3)}\rho_Q-p_Q\rho_Q^{(3)}$ becomes as
\begin{eqnarray}
&=& \left[18\frac{\Delta \sqrt{\frac{8\pi G}{3}\rho_c}}{1-\frac{\Delta}{2}}\frac{V''}{V}-2\frac{\Delta \sqrt{\rho_Q\Delta}}{1-\frac{\Delta}{2}}\frac{V'''}{V}+\frac{24\pi G \left(\frac{2}{3}-\Delta \right)\sqrt{\rho_Q\Delta}}{1-\frac{\Delta}{2}}\frac{V'}{V}-\frac{64\pi G \sqrt{\rho_Q\Delta}}{\Omega_Q\left(1-\frac{\Delta}{2}\right)}\frac{V'}{V} \right. \nonumber \\
&+& \left. 8\sqrt{\rho_Q\Delta}\frac{V'}{V}\frac{V''}{V}-\sqrt{96\pi G\rho_c}\left(\frac{V'}{V}\right)^2\right] \frac{1}{2}\Delta \rho_Q^3\left(1-\frac{\Delta}{2}\right)^2 \nonumber \\
&+& \left[-\frac{96\pi G\sqrt{\frac{8\pi G}{3}\rho_c}\Delta^2}{\left(1-\frac{\Delta}{2}\right)^2}-\frac{48\pi G\Delta \sqrt{\rho_Q\Delta}}{1-\frac{\Delta}{2}}\frac{V'}{V}-\frac{24\pi G\Delta \sqrt{\frac{8\pi G}{3}\rho_c}\left(\frac{2}{3}-\Delta \right)}{\left(1-\frac{\Delta}{2}\right)^2} \right. \nonumber \\
&+& \left. \frac{504\pi G\left(\frac{2}{3}-\Delta \right)\Delta\sqrt{\frac{8\pi G}{3}\rho_c}}{\left(1-\frac{\Delta}{2}\right)^2}-\frac{896\pi G\sqrt{\frac{8\pi G}{3}\rho_c}\Delta}{\Omega_Q\left(1-\frac{\Delta}{2}\right)^2}+30\frac{\Delta \sqrt{\frac{8\pi G}{3}\rho_c}}{1-\frac{\Delta}{2}}\frac{V''}{V}-2\frac{\Delta \sqrt{\rho_Q\Delta}}{1-\frac{\Delta}{2}}\frac{V'''}{V} \right. \nonumber \\
&+& \left. \frac{120\pi G \left(\frac{2}{3}-\Delta \right)\sqrt{\rho_Q\Delta}}{\left(1-\frac{\Delta}{2}\right)}\frac{V'}{V}-\frac{416\pi G \sqrt{\rho_Q\Delta}}{\Omega_Q\left(1-\frac{\Delta}{2}\right)}\frac{V'}{V}+8\sqrt{\rho_Q\Delta}\frac{V'}{V}\frac{V''}{V} \right. \nonumber \\
&-& \left. 18\sqrt{\frac{8\pi G}{3}\rho_c}\left(\frac{V'}{V}\right)^2\right]\rho_Q^3\left(1-\frac{\Delta}{2}\right)^3 .
\end{eqnarray}
To combine the first and second term, we must take factor out the first term by $\times \frac{\Delta}{1-\frac{\Delta}{2}}$ and the second term by $\times \frac{\Delta^2}{2\left(1-\frac{\Delta}{2}\right)^2}$ and arrange as
\begin{eqnarray}
&=& \frac{1}{2}\Delta^2\rho_Q^3\left(1-\frac{\Delta}{2}\right) \times \left[-\frac{4}{\Delta}\sqrt{\rho_Q\Delta} \frac{V'''}{V}+\frac{12}{\Delta}\sqrt{\frac{8\pi G}{3}\rho_c}(-\Delta+5)\frac{V''}{V} \right. \nonumber \\
&+& \left. \frac{8\pi G\sqrt{\rho_Q\Delta}}{\Delta}\left \{-10+3\Delta+\frac{44\Delta-104+30\Omega_Q\left(\frac{2}{3}-\Delta \right)\left(1-\frac{\Delta}{2}\right)}{\Delta\Omega_Q} \right \} \frac{V'}{V} \right. \nonumber \\
&+& \left. \frac{16\sqrt{\rho_Q\Delta}\left(1-\frac{\Delta}{2}\right)}{\Delta^2}\frac{V'}{V}\frac{V''}{V}-\frac{12\sqrt{\frac{8\pi G \rho_c}{3}}\left(1-\frac{\Delta}{2}\right)}{\Delta}\left(-1+\frac{3}{\Delta}\right)\left(\frac{V'}{V}\right)^2 \right. \nonumber \\
&+& \left. 16\pi G\sqrt{\frac{8\pi G}{3}\rho_c}\left \{ -72+\frac{1}{\Delta\Omega_Q}(40\Omega_Q-112) \right \} \right] .
\end{eqnarray}
Subsequently we concentrate the second and third terms, and sort out as 
\begin{eqnarray*}
\hspace{-10cm}\ddot{p_Q}\dot{\rho_Q}-\dot{p_Q} \ddot{\rho_Q} 
\end{eqnarray*}
\begin{eqnarray}
&=& \left[\left(-3\frac{\ddot{a}}{a}+21H^2-2V''\right)\dot{Q}^2 + 12H\dot{Q}V' +2V'^2\right](-3H\dot{Q}^2) \nonumber \\
&+& (3H\dot{Q}^2+2V'\dot{Q})\left[ \left(-3\frac{\ddot{a}}{a} +21H^2\right)\dot{Q}^2+6H\dot{Q}V'\right] \nonumber \\
\nonumber \\
&=& -3H\dot{Q}^2V^2\left[-2\frac{V''}{V}\frac{\dot{Q}^2}{V}+6\frac{H\dot{Q}}{V}\frac{V'}{V}+2\left(\frac{V'}{V}\right)^2\right]+2V'V\dot{Q}\left[ \left(-3\frac{\ddot{a}}{a} +21H^2\right)\frac{\dot{Q}^2}{V}+6H\dot{Q}\frac{V'}{V}\right] \nonumber \\
\nonumber \\
&=& 6H\dot{Q}^4V\frac{V''}{V}+\left( 24H^2\dot{Q}^3V-6\frac{\ddot{a}}{a}V\dot{Q}^3 \right)\frac{V'}{V} +6H\dot{Q}^2V^2\left(\frac{V'}{V}\right)^2 .  \label{aaa}
\end{eqnarray}
The necessary elements are displayed;
\begin{eqnarray}
H\dot{Q}^4V &=& \sqrt{\frac{8\pi G}{3}\rho_c}\rho_Q^3\Delta^2 \left(1-\frac{\Delta}{2}\right) \\
H^2\dot{Q}^3V &=& \frac{8\pi G}{3}\rho_c\rho_Q^2\Delta \sqrt{\rho_Q\Delta} \left(1-\frac{\Delta}{2}\right) \\
\frac{\ddot{a}}{a}V\dot{Q}^3 &=& 4\pi G\rho_Q^3\sqrt{\rho_Q\Delta}\left(\frac{2}{3}-\Delta \right)\left(1-\frac{\Delta}{2}\right)\Delta \\
H\dot{Q}^2V^2 &=& \sqrt{\frac{8\pi G}{3}\rho_c}\rho_Q^3\Delta\left(1-\frac{\Delta}{2}\right)^2.
\end{eqnarray}
Then Eq. (\ref {aaa}) becomes 
\begin{eqnarray}
&=& 6\sqrt{\frac{8\pi G}{3}\rho_c}\rho_Q^3\Delta^2 \left(1-\frac{\Delta}{2}\right)\frac{V''}{V}+\left \{ 64\pi G\rho_c\rho_Q^2\Delta \sqrt{\rho_Q\Delta} \left(1-\frac{\Delta}{2}\right) \right. \nonumber \\
&-& \left. 24\pi G\rho_Q^3\sqrt{\rho_Q\Delta}\left(\frac{2}{3}-\Delta \right)\left(1-\frac{\Delta}{2}\right)\Delta \right \}\frac{V'}{V}+6\sqrt{\frac{8\pi G}{3}\rho_c}\rho_Q^3\Delta\left(1-\frac{\Delta}{2}\right)^2\left(\frac{V'}{V}\right)^2 .\nonumber 
\end{eqnarray}
So we must take factor out the whole equation by $\frac{\Delta^2\rho_Q^3\left(1-\frac{\Delta}{2}\right)}{2}$ and arrange as
\begin{eqnarray}
\frac{1}{2}\Delta^2\rho_Q^3\left(1-\frac{\Delta}{2}\right) & \times & \left[12\sqrt{\frac{8\pi G}{3}\rho_c}\frac{V''}{V}+\left \{ \frac{128\pi G\sqrt{\rho_Q\Delta}}{\Delta\Omega_Q}-\frac{48\pi G\left(\frac{2}{3}-\Delta \right)\sqrt{\rho_Q\Delta}}{\Delta}\right\}\frac{V'}{V} \right. \nonumber \\
&+& \left. \frac{12\sqrt{\frac{8\pi G}{3}\rho_c}\left(1-\frac{\Delta}{2}\right)}{\Delta}\left(\frac{V'}{V}\right)^2 \right] .
\end{eqnarray}
Then the first Equation of $p_Q^{(3)}\rho_Q+\ddot{p_Q}\dot{\rho_Q}-\dot{p_Q} \ddot{\rho_Q}-p_Q\rho_Q^{(3)}$ becomes as
\begin{eqnarray}
&=& \frac{1}{2}\Delta^2\rho_Q^3\left(1-\frac{\Delta}{2}\right)\left[-\frac{4}{\Delta}\sqrt{\rho_Q\Delta} \frac{V'''}{V}+\frac{60}{\Delta}\sqrt{\frac{8\pi G}{3}\rho_c}\frac{V''}{V} \right. \nonumber \\
&+& \left. \frac{8\pi G\sqrt{\rho_Q\Delta}}{\Delta}\left \{-14+9\Delta+\frac{60\Delta-104+30\Omega_Q\left(\frac{2}{3}-\Delta \right)\left(1-\frac{\Delta}{2}\right)}{\Delta\Omega_Q} \right \} \frac{V'}{V} \right. \nonumber \\
&+& \left. \frac{16\sqrt{\rho_Q\Delta}\left(1-\frac{\Delta}{2}\right)}{\Delta^2}\frac{V'}{V}\frac{V''}{V}+\frac{12\sqrt{\frac{8\pi G}{3}\rho_c}\left(1-\frac{\Delta}{2}\right)}{\Delta}\left(2-\frac{3}{\Delta}\right)\left(\frac{V'}{V}\right)^2 \right. \nonumber \\
&+& \left. 16\pi G\sqrt{\frac{8\pi G}{3}\rho_c}\left \{ -72+\frac{1}{\Delta\Omega_Q}(40\Omega_Q-112) \right \} \right] .
\end{eqnarray}

To put together the above whole calculations, the first term of Eq. (\ref{d3w/da3Appe}) becomes
\begin{eqnarray}
&  \ & \left \{ \frac{1}{2}\Delta^2\rho_Q^3\left(1-\frac{\Delta}{2}\right) \times a\sqrt{\frac{8\pi G}{3}\rho_c} \rho_Q^2 \left[-\frac{4}{\Delta}\sqrt{\rho_Q\Delta} \frac{V'''}{V}+\frac{60}{\Delta}\sqrt{\frac{8\pi G}{3}\rho_c}\frac{V''}{V} \right. \right. \nonumber \\
&+& \left. \left.  \frac{8\pi G\sqrt{\rho_Q\Delta}}{\Delta}\left \{-14+9\Delta+\frac{60\Delta-104+30\Omega_Q\left(\frac{2}{3}-\Delta \right)\left(1-\frac{\Delta}{2}\right)}{\Delta\Omega_Q} \right \} \frac{V'}{V} \right. \right. \nonumber \\
&+& \left. \left. 16\frac{\sqrt{\rho_Q\Delta}\left(1-\frac{\Delta}{2}\right)}{\Delta}\frac{V'}{V}\frac{V''}{V}+\frac{12\sqrt{\frac{8\pi G}{3}\rho_c}\left(1-\frac{\Delta}{2}\right)}{\Delta}\left(2-\frac{3}{\Delta}\right)\left(\frac{V'}{V}\right)^2 \right. \right. \nonumber \\
&+& \left. \left. 16\pi G\sqrt{\frac{8\pi G}{3}\rho_c}\left \{ -72+\frac{1}{\Delta\Omega_Q}(40\Omega_Q-112) \right \} \right] \right. \nonumber \\
&+& \left. \frac{8\pi Ga \rho_Q^5 \rho_c^{\frac{1}{2}}}{\Omega_Q}\left(1-\frac{\Delta}{2}\right) \left[\sqrt{24\pi G}\Delta + \frac{V'}{V}\sqrt{\Delta \Omega_Q} \right] \left[H\Omega_Q\left(\frac{2}{3}-\Delta\right)(1-18\Delta) \right. \right. \nonumber \\
&+& \left. \left. 4H\Delta(\Omega_Q+3\Delta+7)+2\frac{V'}{V}\left(1-\frac{\Delta}{2}\right)\sqrt{\Delta\rho_Q}(\Omega_Q+4)\right] \right \} \times a\sqrt{\frac{8\pi G}{3}\rho_c} \rho_Q  . \nonumber \\
\end{eqnarray}
Then we put togather the above equation by  Eq. (\ref{ThirdDeriSec}) times by $-1$  and facotorize by $\frac{1}{2}a\rho_Q^5\left(1-\frac{\Delta}{2}\right)\sqrt{\frac{8\pi G}{3}\rho_c}$.  
Then Eq. (\ref{d3w/da3Appe}) becomes to Eq. (\ref{d3w/da3}).

\vspace{1cm}
\hspace{-0.5cm} {\large{\bf References}}


\end{document}